\documentclass[aps,amsmath,twocolumn,amssymb,floatfixng,showpacs,
superscriptaddress,footinbib]{revtex4-2}

\usepackage{bm}
\usepackage{graphicx}
\usepackage{graphics}
\usepackage{mathtools}
\usepackage{amsmath}
\usepackage{amsfonts}
\usepackage{amssymb}
\usepackage{epstopdf}
\usepackage{hyperref}
\usepackage{hyperref}
\usepackage{booktabs}
\usepackage{tabularx}
\hypersetup{
    colorlinks,%
    citecolor=blue,%
    linkcolor=blue,%
    urlcolor=blue
}
\usepackage[normalem]{ulem}
\usepackage{tikz}\usetikzlibrary{petri}

\newcommand{\be}   {\begin{equation}}
\newcommand{\ee}   {\end{equation}}
\newcommand{\ba}   {\begin{eqnarray}}
\newcommand{\ea}   {\end{eqnarray}}

\begin{document}

\title{Interlayer interactions in $\text{La}_3\text{Ni}_2\text{O}_7$ under pressure: from  $s^{\pm}$ to $d_{xy}$-wave superconductivity}
\author{Lauro B. Braz}
\affiliation{
Instituto de F\'{\i}sica, Universidade de S\~ao Paulo, Rua do Mat\~ao 1371, S\~ao Paulo, S\~ao Paulo 05508-090, Brazil
}
\author{George B. Martins}
\affiliation{Instituto de F\'isica, Universidade Federal de Uberl\^andia, 
Uberl\^andia, Minas Gerais 38400-902, Brazil.
}
\author{Luis G.~G.~V. Dias da Silva}
\affiliation{
Instituto de F\'{\i}sica, Universidade de S\~ao Paulo, Rua do Mat\~ao 1371, S\~ao Paulo, S\~ao Paulo 05508-090, Brazil
}

\date{ \today }

\begin{abstract}
We investigate the role of \emph{interlayer} interaction terms in the competition between different superconducting gap symmetries in the bilayer nickelate $\text{La}_3\text{Ni}_2\text{O}_7$ under high pressure.  We study a two-layer, two-orbital electron model that encompasses both intra- and interlayer Coulomb interaction terms within the matrix random-phase approximation. We find that interlayer interactions favor a $d_{xy}$-wave superconducting pairing symmetry over the $s^{\pm}$-wave symmetry, which has been found to prevail when interlayer interactions are disregarded. Moreover, our findings indicate that interlayer interactions enhance the interorbital pairing, incorporating contributions from all three electron pockets, arising from both $d_{3z^2-r^2}$ and $d_{x^2-y^2}$ orbital character, resulting in nodes within the gap function (not present in the $s^{\pm}$-wave state) and consequently favoring the $d_{xy}$-wave pairing.
\end{abstract}

\maketitle

\section{Introduction}
\label{sec:Intro}

The recent discovery of high-temperature superconductivity in the Ruddlesden-Popper bilayer nickelate compound $\text{La}_3\text{Ni}_2\text{O}_7$ (LNO327) under high pressure \cite{sunSignaturesSuperconductivity802023,
wangPressureInducedSuperconductivityPolycrystalline2024,Li:NickelateMeissner:arXiv:2024} has sparked intense research on this new family of high-$T_c$ materials  \cite{WangReview:ChinesePhysicsLetters:077402:2024}. Despite the rapid progress in the last two years, including recent reports of superconductivity on thin films at ambient pressure \cite{Ko:Nature::2024},  several fundamental questions remain on the nature of the underlying pairing mechanism for superconductivity in these compounds. 

For example, there has been an intense debate as to which low-energy \emph{minimal} model to describe the superconducting phase.
Density Functional Theory calculations \cite{luoBilayerTwoOrbitalModel2023, zhangElectronicStructureDimer2023, zhangStructuralPhaseTransition2024, lechermannElectronicCorrelationsSuperconducting2023} reveal that multiple bands,  mainly formed by Ni-centered in-plane $d_{x^2-y^2}$ and out-of-plane $d_{3z^2-r^2}$ orbitals, along with oxygen-centered $p$ orbitals  cross the Fermi level, forming two electron pockets ($\alpha$, $\beta$) with mixed orbital character and one hole pocket ($\gamma$) primarily associated with the Ni $d_{3z^2-r^2}$ orbital \cite{luoBilayerTwoOrbitalModel2023}. These bands can be described by a two-layer, two-orbital minimal tight-binding model  \cite{luoBilayerTwoOrbitalModel2023,zhangElectronicStructureDimer2023,lechermannElectronicCorrelationsSuperconducting2023}, which has been investigated with interlayer interactions within 
a localized electron picture~\cite{luoHighTCSuperconductivityLa3Ni2O72024} and without 
interlayer interactions in an itinerant electron picture 
\cite{yangPossiblewaveSuperconductivity2023,
liuWavePairingDestructive2023,zhangStructuralPhaseTransition2024,lechermannElectronicCorrelationsSuperconducting2023,lechermannElectronicInstabilityLayer2024}. 

Using these models,  some RPA calculations suggest that $s^{\pm}$-wave pairing is the dominant superconducting instability \cite{yangPossiblewaveSuperconductivity2023,quBilayerModel2024,chenOrbitalselectiveSuperconductivityPressurized2024,yangPossiblewaveSuperconductivity2023,liuWavePairingDestructive2023,sakakibaraPossibleHigh$T_c$2024,zhangStructuralPhaseTransition2024}, which can be driven by intra-orbital ($\pi$, 0) scattering between the $\beta$ and the $\gamma$ pockets \cite{zhangStructuralPhaseTransition2024}, although  the $d$-wave pairing channel is found to be close in energy to the $s$-wave pairing one \cite{liuWavePairingDestructive2023,zhangStructuralPhaseTransition2024}. 

By contrast, several works have brought up the possibility that $d$-wave symmetry is dominant in some regimes. These range from  $d$-$p$ hybridized single-orbital models showing cuprate-like $d_{x^2-y^2}$ symmetry \cite{fanSuperconductivityNickelateCuprate2024} to two-orbital models based on Ni-$e_g$ DFT bands \cite{lechermannElectronicCorrelationsSuperconducting2023} as well as the differences in the spin response in these cases \cite{botzelTheoryMagneticExcitations2024}. Moreover, a recent paper \cite{Xia:NatureCommunications:1054:2025} reports DFT+RPA calculations showing that the leading superconducting instability has $d_{xy}$ symmetry, although a slight increase in the strength of the Ni-$e_g$ crystal field splitting can change the pairing symmetry to $s^{\pm}$.

On the experimental side, recent explorations of the ambient pressure
properties of LNO327  give  indications that \emph{interlayer} interactions can be relevant for the superconducting phase at high pressures. Neutron scattering experiments in polycrystalline samples show evidence of spin excitations in the inelastic channel, which are consistent with strong interlayer exchange couplings \cite{xieStrongInterlayerMagnetic2024} while X-ray absorption measurements \cite{Chen:NatureCommunications:9597:2024} revealed  dispersive magnons and spin-density wave order patterns which are also consistent with interlayer magnetic exchange being about an order of magnitude larger than intralayer ones. 
In addition, recent RIXS experiments in LNO327 thin films samples show evidence of  an strain-enhanced interlayer exchange interaction coupling \cite{Zhong_arXiv.org2502.03178__2025}.

In this work, we investigate the role of the interlayer Coulomb interaction terms in the competition between different superconducting gap symmetries in LNO327 under high pressure. We consider a two-layer, two-orbital electron model that includes both intra- and interlayer Coulomb interaction terms.  
Within the weak- to intermediate-coupling matrix random-phase approximation (mRPA) \cite{graserDegeneracySeveralPairing2009,kemperSensitivitySuperconductingState2010,altmeyerRoleVertexCorrections2016,Martins:Phys.Rev.B:081102:2013,brazChargeSpinFluctuations2024}, 
we find that interlayer interactions 
favor a $d_{xy}$-wave superconductivity over the $s^{\pm}$ state, which is known to dominate when such interlayer interactions are neglected \cite{yangPossiblewaveSuperconductivity2023,zhangStructuralPhaseTransition2024,quBilayerModel2024}. 
Our mRPA calculations reveal that interlayer interactions enhance interorbital pairing involving contributions from \emph{all three} pockets, creating nodes in the gap function (absent in the $s^{\pm}$-wave state) and thus favoring the $d_{xy}$-wave pairing.

This paper is organized as follows: in Section \ref{sec:Model}, we present the interacting model for bilyer LNO327, including the band structure and both the onsite and interlayer interaction terms. We review the matrix-RPA methodology and describe the paring gap calculation in Section \ref{sec:mRPA}. Our results detailing the $s^{\pm}$-wave to $d_{xy}$-wave symmetry transition as the interlayer interaction increases are presented in Sec.~\ref{sec:pairingsymmetries}, along with a discussion on the role of the charge susceptibility and the orbital contributions. Finally, our concluding remarks are presented in Section \ref{sec:Conclusions}.


\section{Model}
\label{sec:Model}

\subsection{Band structure}
\label{sec:BandStructure}

\begin{figure}[t]
\begin{center}
\includegraphics[width=1\linewidth]{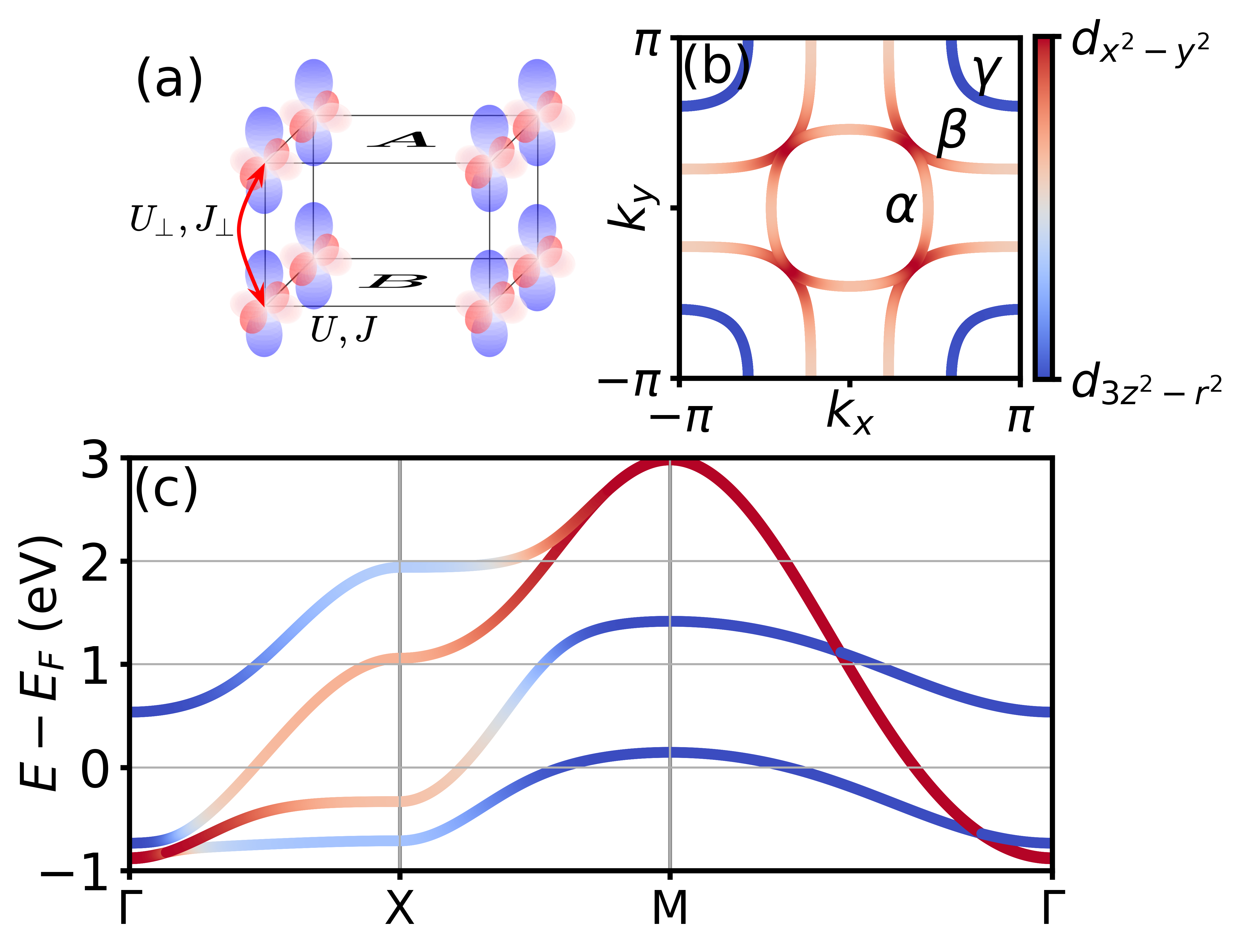}
\end{center}
\vspace{-17pt}
\caption{ Schematic representation of the bilayer $\text{La}_3\text{Ni}_2\text{O}_7$ model (a) with onsite $U,J$ and interlayer $U_\perp,J_\perp$ Coulomb interaction and Hund's exchange coupling  strengths. FS (b) and band structure (c) for the two-orbital $d_{3z^2-r^2}$ (blue color) and $d_{x^2-y^2}$ (red color) low-energy model.
}
\label{fig:model_system}
\end{figure}

In order to describe the (non-interacting) band structure of the system, we consider the four-band low-energy model of Refs. \cite{luoBilayerTwoOrbitalModel2023,liuWavePairingDestructive2023}. This tight-binding model involves two orbitals ($d_{x^2-y^2}$ and $d_{3z^2-r^2}$) per Ni atom in a unit cell with two Ni atoms (one for each layer) [as depicted in Fig.~\ref{fig:model_system}(a)] and hoppings up to second nearest neighbors. 
The non-interacting Hamiltonian reads
\begin{equation}
    \begin{split}
        H_{0}	&=\sum_{\boldsymbol{k}\sigma p \ell}(T_{\boldsymbol{k}}^{p}-\mu)c_{\boldsymbol{k}p \ell \sigma}^{\dagger}c_{\boldsymbol{k} p \ell \sigma} \\
	&+\sum_{\boldsymbol{k}\sigma p}t_{\perp}^{p}(c_{\boldsymbol{k}pA\sigma}^{\dagger}c_{\boldsymbol{k}pB\sigma}+\text{h.c.}) \\
    &+\sum_{\boldsymbol{k}\sigma(p\neq s) \ell }V_{\boldsymbol{k}}c_{\boldsymbol{k}p \ell \sigma}^{\dagger}c_{\boldsymbol{k}s \ell \sigma} \\
	&+\sum_{\boldsymbol{k}\sigma(p\neq s)}V_{\boldsymbol{k}}^{\prime}(c_{\boldsymbol{k}pA\sigma}^{\dagger}c_{\boldsymbol{k}sB\sigma}+\text{h.c.}),
    \end{split}
\label{eq:H0}
\end{equation}
where $c^{\dagger}_{\boldsymbol{k}p \ell \sigma}$ creates an electron with spin $\sigma$, in the orbital $p=\{x,z\}$ corresponding to $\{d_{x^2-y^2},d_{3z^2-r^2}\}$, sublattice $\ell=\{A,B\}$, and momentum $\boldsymbol{k}$, $\mu$ is the chemical potential, and h.c. shortcuts Hermitian conjugate.
In the following, we use the model parameters of Ref. \cite{luoBilayerTwoOrbitalModel2023}, which includes on-site energies ($\epsilon^{p}$) and up to second-neighbor hoppings such that
\begin{equation}
    \begin{split}
        T_{\boldsymbol{k}}^{p}&	=2t_{1}^{p}(\cos k_{x}+\cos k_{y})+4t_{2}^{p}\cos k_{x}\cos k_{y}+\epsilon^{p} \; , \\
V_{\boldsymbol{k}}	&=2t_{3}^{xz}(\cos k_{x}-\cos k_{y}) \; ,\\
V_{\boldsymbol{k}}^{\prime}&=2t_{4}^{xz}(\cos k_{x}-\cos k_{y}) \; .
    \end{split}
\label{eq:H0_terms}
\end{equation}

For a valence of 2.5 electrons per Ni atom \cite{Cai:arXiv:2412.18343:2024}, the resulting Fermi surface (FS) and band structure are shown in Fig.~\ref{fig:model_system}(b) and (c), respectively, together with the respective Ni orbital contributions.

The FS shows electron pockets $\alpha$ and $\beta$ with a strong contribution from the  $d_{x^2-y^2}$ orbital (red),  consistent with ARPES results at ambient pressure \cite{yangOrbitaldependentElectronCorrelation2024,liElectronicCorrelationPseudogapBehavior2024}.
The $\gamma$ hole pocket at $(\pi,\pi)$, however, is dominated by the $d_{3z^2-r^2}$ orbital (blue) and its presence has been linked to the onset of superconductivity if LNO327 and similar materials \cite{zhangStructuralPhaseTransition2024,Zhang:Phys.Rev.B:045151:2024,Zhang:Phys.Rev.Lett.:136001:2024}. We notice that some of the bands show 
mixed orbital regions (nearly white) along the $\boldsymbol{\Gamma}-\boldsymbol{X}$ and $\boldsymbol{X}-\boldsymbol{M}$ high-symmetry paths [Fig.~\ref{fig:model_system}(c)].

\subsection{Onsite and interlayer interactions}
\label{sec:interactions}

Our model Hamiltonian $H=H_0+H_I$ involves both the non-interacting part described by $H_0$ as well as an interacting term ($H_I$) that includes both onsite ($U,J$) and interlayer ($U_\perp,J_\perp$) Coulomb and Hund's coupling interaction terms, as represented in Fig.~\ref{fig:model_system}(a). 
We remark that $J,J_\perp$ are the Hund's couplings and differ from the effective exchange interaction
 $\tilde{J},\tilde{J}_{\perp}$ of strong-coupling models \cite{quBilayerModel2024,
chenOrbitalselectiveSuperconductivityPressurized2024}.

The interacting Hamiltonian for the two-orbital itinerant electron model is given by the onsite plus interlayer parts, $H_I=H_{\text{onsite}}+H_{\text{inter}}$, where the onsite term is given by
\begin{equation}
    \begin{split}
        H_{\text{onsite}} &= U\sum_{ij ,p \ell }n_{ip \ell \uparrow}n_{ip \ell \downarrow} \\
        & + (U'-J/2)\sum_{i\ell,p<q}n_{ip\ell}n_{iq\ell} \\
        & - 2J\sum_{i\ell,p<q} \boldsymbol{S}_{ip\ell}\cdot\boldsymbol{S}_{iq\ell} \\
        & + J^{\prime} \sum_{i\ell,p<q}\left( c^{\dagger}_{ip\ell\uparrow}c^{\dagger}_{ip\ell\downarrow}c_{iq\ell\downarrow}c_{iq\ell\uparrow} +\text{h.c.}\right)
    \end{split}
\label{eq:H_on}
\end{equation}
and the interlayer term reads 
\begin{equation}
    \begin{split}
        H_{\text{inter}} &= U_{\perp}\sum_{ij,p\sigma\bar{\sigma}}n_{ipA\sigma}n_{ipB\bar{\sigma}} \\
        & + (U_{\perp}'-J_{\perp}/2)\sum_{i,p<q}n_{ipA}n_{iqB} \\
        & - 2J_{\perp}\sum_{i,p<q} \boldsymbol{S}_{ipA}\cdot\boldsymbol{S}_{iqB} \\
        & + J^{\prime}_{\perp}\sum_{i,p<q}\left( c^{\dagger}_{ipA\uparrow}c^{\dagger}_{ipB\downarrow}c_{iqB\downarrow}c_{iqA\uparrow} +\text{h.c.}\right).
    \end{split}
\label{eq:H_inter}
\end{equation}

In the above, $c^{\dagger}_{ip \ell \sigma}$ creates an electron with spin $\sigma$, in orbital $p$, layer $\ell=A,B$, at the (square) lattice site $i$.
In addition, we use a shortcut notation for the density operator, $n_{ip \ell}=\sum_{\sigma} n_{ip \ell \sigma}$.
In the present model, $U(U_\perp)$, $U'(U_\perp')$, $J(J_\perp)$, and $J^{\prime}(J^{\prime}_{\perp})$ are the onsite (intersite) Hubbard, interorbital Hubbard, Hund's exchange coupling and pair-hopping interaction strengths, respectively. In the following, we assume local spin rotational invariance such that $J^{\prime}\!=\!J (J^{\prime}_{\perp}\!=\!J_\perp)$ \cite{graserDegeneracySeveralPairing2009,kemperSensitivitySuperconductingState2010}.

The interacting part of the Hamiltonian, $H_I$, can conveniently be rewritten in momentum space as
\begin{equation}
    \begin{split}
        H_I & = \sum_{\boldsymbol{q}} \sum_{pqst \ell \ell'} (U_c)^{p\ell q \ell }_{s \ell' t \ell'} n_{p\ell q \ell}(-\boldsymbol{q})n_{s \ell' t \ell'}(\boldsymbol{q}) \\
        & + \sum_{\boldsymbol{q}} \sum_{pqst \ell \ell'} (U_s)^{p\ell q \ell}_{s\ell' t\ell' } \boldsymbol{S}_{p\ell q\ell}(-\boldsymbol{q})\cdot\boldsymbol{S}_{s\ell' t \ell'}(\boldsymbol{q}),
    \end{split}
\label{eq:H_density}
\end{equation}
where $n_{pAqA}(\boldsymbol{q})=\sum_{\boldsymbol{k}\sigma}c^{\dagger}_{(\boldsymbol{k}+\boldsymbol{q})pA\sigma}c_{\boldsymbol{k}qA\sigma}$ is the charge density and $\boldsymbol{S}_{pAqA}(\boldsymbol{q})=\sum_{\boldsymbol{k}\sigma\bar{\sigma}}c^{\dagger}_{(\boldsymbol{k}+\boldsymbol{q})pA\sigma}\boldsymbol{\sigma}_{\sigma\bar{\sigma}}c_{\boldsymbol{k}qA\bar{\sigma}}$ the spin density for layer $A$ (similarly for layer $B$), with $\boldsymbol{\sigma}_{\sigma\bar{\sigma}}$ being the Pauli vector. 
The matrix elements of the charge and spin interaction matrices which connect Eq.~\eqref{eq:H_density} with Eqs.~\eqref{eq:H_on} and \eqref{eq:H_inter} are given by
\begin{equation}
    \begin{split}
        & (U_s)^{pApA}_{pApA}=U, \; (U_s)^{pApA}_{qAqA}=U', \\
        & (U_s)^{pAqA}_{pAqA}=(U_s)^{pAqA}_{qApA}=J, \\
        & (U_c)^{pApA}_{pApA}=U, \; (U_c)^{pApA}_{qAqA}=-U'+2J, \\
        & (U_c)^{pAqA}_{qApA}=J, \; (U_c)^{pAqA}_{pAqA}=-J+2U', \\
        & (U_c)^{pApA}_{pBpB}=2U_{\perp}, \; (U_c)^{pApA}_{qBqB}=2U'_{\perp}, \\
        & (U_c)^{pAqA}_{pBqB}=(U_c)^{pAqA}_{qBpB}=2J_{\perp}, \\
    \end{split}
\label{eq:U_matrices}
\end{equation}
where $A,B$ denote the respective layer index \footnote{The remaining matrix elements can be trivially obtained by exchanging $A \leftrightarrow B$}.
For simplicity, we assume spin rotational invariance such that the standard relationships $U'=U-2J$ and $U'_{\perp}=U_{\perp}-2J_{\perp}$ hold and we  
set $J=U/4$ and $J_{\perp}=U_{\perp}/4$, which is consistent with previous studies \cite{zhangElectronicStructureDimer2023}, although we have verified that this choice does not qualitatively alter our main results  (see Appendix \ref{app:J_Jperp}). Notice that the interlayer interaction terms $(U_{\perp},U'_{\perp})$ appear only in the \emph{charge} interaction matrices.

\section{Matrix RPA method}
\label{sec:mRPA}

\subsection{Spin and charge susceptibilities}
\label{sec:spinchargesuscep}

For completeness, we begin this Section by briefly reviewing the RPA-enhanced spin and charge susceptibilities for the
multiorbital model of Eq.~\eqref{eq:H0}.  Following Ref. \cite{graserDegeneracySeveralPairing2009}, the bare spin susceptibility is calculated from the spin-spin correlation function as
\begin{equation}
    \begin{split}
        [\hat{\chi}]^{p\ell q\ell'}_{s\xi t\xi'}(\boldsymbol{q},i\nu_m) &= \int^{1/T}_0d\tau e^{i\nu_m\tau}\\
        &\times\langle T_{\tau}\boldsymbol{S}_{p\ell q\ell'}(-\boldsymbol{q},\tau)\cdot\boldsymbol{S}_{s\xi t\xi'}(\boldsymbol{q},0)\rangle,
    \end{split}
\end{equation}
where $i\nu_m$ denotes Matsubara bosonic frequencies, $\tau$ is the imaginary time, $T$ is the temperature, and $T_{\tau}$ is the time-ordering operator. For the non-interacting model, the spin and charge susceptibilities are equivalent \cite{graserDegeneracySeveralPairing2009} and can be written in terms of the non-interacting Green's functions as
\begin{equation}
    \begin{split}
        [\hat{\chi}]^{p\ell q\ell'}_{s\xi t\xi'}(\boldsymbol{q},i\nu_m)=&-\frac{T}{N}\sum_{\boldsymbol{k}i\omega_n}G_{s\xi p\ell}(\boldsymbol{k},i\omega_n)\\
        &\times G_{q\ell't\xi'}(\boldsymbol{k}+\boldsymbol{q},i\omega_n+i\nu_m),
    \end{split}
    \label{eq:BareSuscep}
\end{equation}
where
\begin{equation}
    G_{s\xi p\ell}(\boldsymbol{k},i\omega_n) = \sum_{\nu_{\boldsymbol{k}}}\frac{\psi^{s\xi}_{\nu_{\boldsymbol{k}}}(\boldsymbol{k})\psi^{p\ell*}_{\nu_{\boldsymbol{k}}}(\boldsymbol{k})}{i\omega_n-E_{\nu_{\boldsymbol{k}}}(\boldsymbol{k})},
\end{equation}
and $\psi^{p \ell}_{\nu_{\boldsymbol{k}}}(\boldsymbol{k})$ is a single-particle eigenfunction of $H_0$ (given by Eq.~\eqref{eq:H0}) related to the $\nu_{\boldsymbol{k}}$-th electronic band with energy $E_{\nu_{\boldsymbol{k}}}(\boldsymbol{k})$,  and written in a basis where $p$ and $\ell$ are good quantum numbers. The Matsubara summation is calculated numerically by using Ozaki's method~\cite{ozakiContinuedFractionRepresentation2007}. In this work, we adopt $T=0.02$ eV.

Within the  mRPA, adding the onsite and interlayer interactions reduces to performing a matrix operation such that the spin and charge susceptibilities become
\begin{align}
    \hat{\chi}_s(\boldsymbol{q},i\nu_m)&=\hat{\chi}(\boldsymbol{q},i\nu_m)[\hat{1}-\hat{U}_s\hat{\chi}(\boldsymbol{q},i\nu_m)]^{-1},\label{eq:chis}\\
    \hat{\chi}_c(\boldsymbol{q},i\nu_m)&=\hat{\chi}(\boldsymbol{q},i\nu_m)[\hat{1}+\hat{U}_c\hat{\chi}(\boldsymbol{q},i\nu_m)]^{-1}. \label{eq:chic}
\end{align}
where $\hat{\chi}(\boldsymbol{q},i\nu_m)$ is the bare susceptibility matrix given by Eq.~\eqref{eq:BareSuscep}. We have checked that the Stoner criterion is not matched for the range of interaction strengths studied in the present work.

\begin{figure}[t]
\begin{center}
\includegraphics[width=1\linewidth]{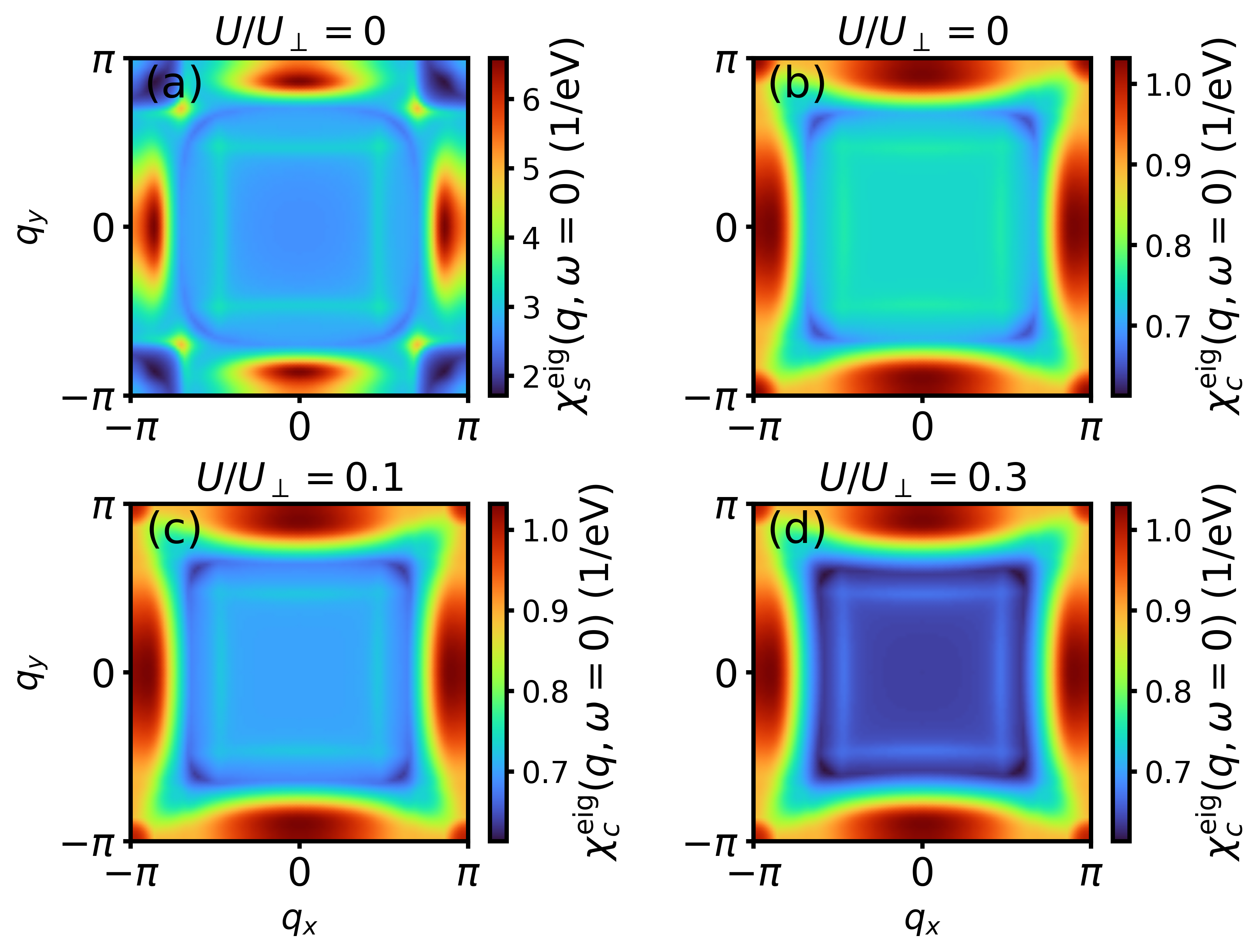}
\end{center}
\vspace{-17pt}
\caption{ Contour plots for the main eigenvalue of the spin (a) and charge (b,c,d) RPA susceptibility matrices for $U=1.16$ eV and different $U_{\perp}/U$ ratios. 
}
\label{fig:Chi}
\end{figure}

It is important to notice that  $\hat{U}_s$ does \emph{not}  depend on $U_{\perp}$ while $\hat{U}_c$ does (see Eq.~\eqref{eq:U_matrices}). As such, only the \emph{charge} susceptibility changes as the interlayer interactions increase. This is illustrated in Figure \ref{fig:Chi}, which shows the main eigenvalue of the spin [$\hat{\chi}_s(\boldsymbol{q},0)$] and charge [$\hat{\chi}_c(\boldsymbol{q},0)$] static susceptibility matrices for different values of $U_{\perp}/U$ for $U=1.16$ eV.
As such, $\hat{\chi}_s(\boldsymbol{q})$ does not change as $U_{\perp}/U$ varies, while the main changes in  $\hat{\chi}_c(\boldsymbol{q})$ are a reduction near the $(\pi,\pi)$ ($\boldsymbol{M}$) point and and overall reduction around the $(0,0)$ ($\boldsymbol{\Gamma}$) point. These features will be further discussed in Section~\ref{sec:chargesuscep}.

\subsection{RPA pairing interaction}
\label{sec:RPApairing}

We now consider the singlet-channel RPA pairing interaction matrix  \cite{graserDegeneracySeveralPairing2009,altmeyerRoleVertexCorrections2016}  for the multi-orbital, multilayer interacting model of Eqs.~\eqref{eq:H_density} and \eqref{eq:U_matrices}
\begin{equation}
    \begin{split}
        \Gamma^{p\ell q\ell'}_{s\xi t\xi'}(\boldsymbol{q}) = \frac{1}{2}\Big[ &-\hat{U}_c\hat{\chi}_c(\boldsymbol{q})\hat{U}_c \\
        &+ 3\hat{U}_s\hat{\chi}_s(\boldsymbol{q})\hat{U}_s + \hat{U}_c + \hat{U}_s \Big]^{t\xi' q\ell' }_{p\ell s\xi}.
    \end{split}
\label{eq:pairing_vertex}
\end{equation}

In the mRPA, the pairing interaction $\Gamma (\boldsymbol{k},\boldsymbol{k}')$ between Cooper pairs at $(\boldsymbol{k},-\boldsymbol{k})$ and $(\boldsymbol{k}',-\boldsymbol{k}')$ is computed from Eq.~\eqref{eq:pairing_vertex} as \cite{graserDegeneracySeveralPairing2009,kemperSensitivitySuperconductingState2010}
\begin{equation}
    \begin{split}
        \Gamma (\boldsymbol{k},\boldsymbol{k}') = \text{Re} &\sum_{\ell \ell' \xi \xi'}\sum_{pqst} \left[\psi^{t\xi'}_{\nu_{-\boldsymbol{k}}}(-\boldsymbol{k})\right]^* \left[\psi^{s\xi *}_{\nu_{\boldsymbol{k}}}(\boldsymbol{k}) \right]^*\\
        &\times\Gamma^{p\ell q\ell'}_{s\xi t\xi'}(\boldsymbol{k}-\boldsymbol{k}')\psi^{p\ell }_{\nu_{\boldsymbol{k}'}}(\boldsymbol{k}')\psi^{q\ell'}_{\nu_{-\boldsymbol{k}'}}(-\boldsymbol{k}') \; .
    \end{split}
\label{eq:pairing_int}
\end{equation}
%

The symmetric part of the pairing interaction $2\bar{\Gamma}(\boldsymbol{k},\boldsymbol{k}')={\Gamma}(\boldsymbol{k},\boldsymbol{k}')+{\Gamma}(\boldsymbol{k},-\boldsymbol{k}')$ is, in turn, used to calculate the gap function $\Delta^\alpha(\boldsymbol{k})$ and the respective pairing strength $\lambda^\alpha$ using the usual integral eigenvalue equation \cite{graserDegeneracySeveralPairing2009}
\begin{equation}
    -\sum_{j}\oint_{\boldsymbol{k'} \in C_{j}}\frac{d\boldsymbol{k}'_{||}}{v_{F}(\boldsymbol{k}')}\frac{1}{(2\pi)^{2}}\bar{\Gamma}(\boldsymbol{k},\boldsymbol{k}')\Delta^{\alpha}(\boldsymbol{k'})=\lambda^{\alpha} \Delta^{\alpha}(\boldsymbol{k}) \; ,
\label{eq:eigenvalues_eq}
\end{equation} 
where the sum runs over the different FS sheets  $C_{j}$ and $v_F(\boldsymbol{k})$ denotes the Fermi speed at momentum $\boldsymbol{k}$.

\begin{figure}[t]
\begin{center}
\includegraphics[width=1\linewidth]{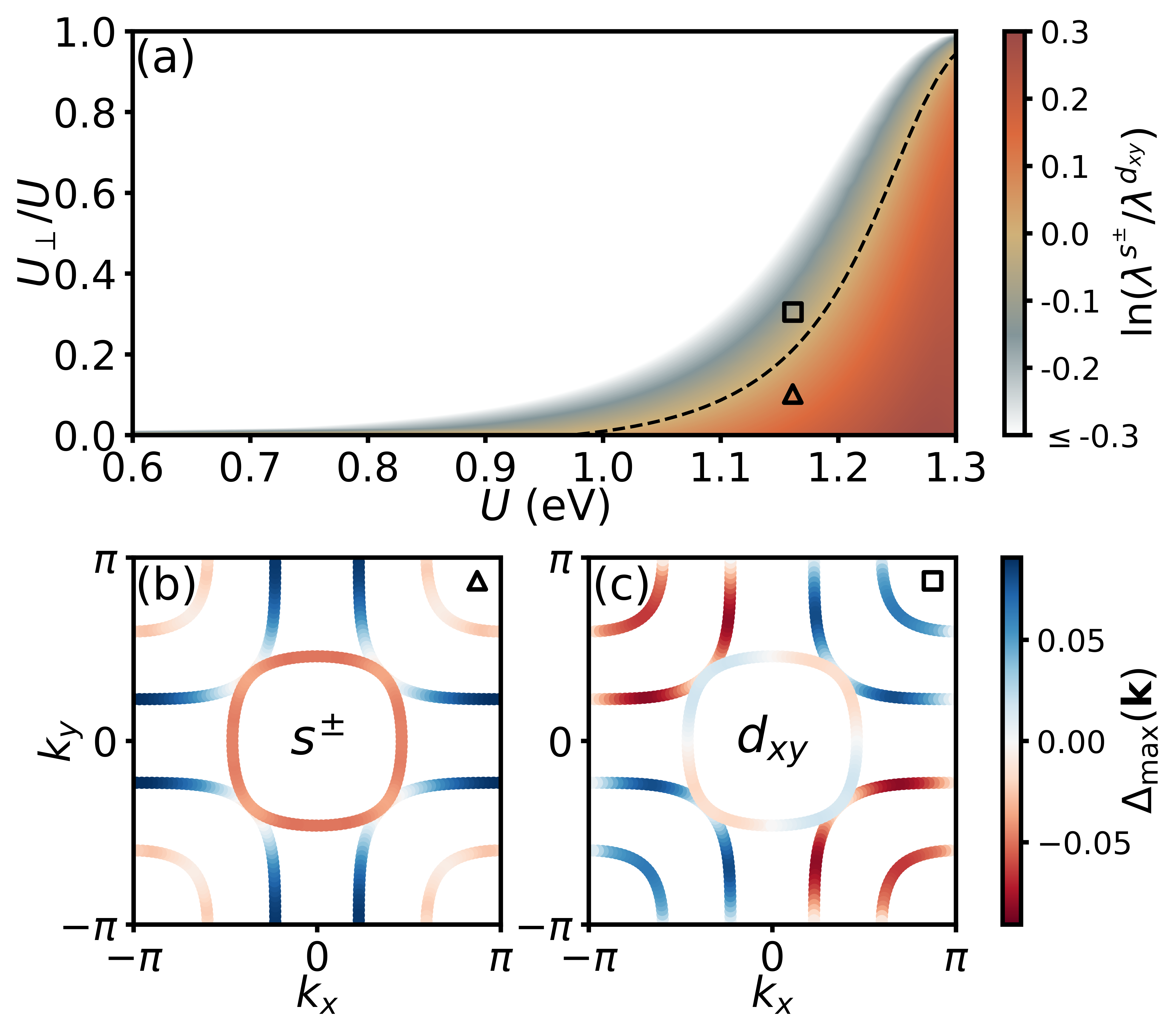}
\end{center}
\vspace{-17pt}
\caption{(a) Phase diagram showing the log of the ratio between the two leading eigenvalues of Eq.~\eqref{eq:eigenvalues_eq}  for given values of $U$ and $U_\perp$. The corresponding leading gap functions $\Delta_{\rm max}(\boldsymbol{k})$ displaying $s^{\pm}$-wave and $d_{xy}$-wave symmetries for $U=1.16$ eV and $U_\perp/U=0.1$ ($\triangle$) and $U_\perp/U=0.3$ ($\square$) are shown in panels (b) and (c), respectively. }
\label{fig:phase_diagram} 
\end{figure}

\section{Superconducting paring symmetries}
\label{sec:pairingsymmetries}

\subsection{Leading solutions of the gap equation}
\label{sec:solutionsgapequation}

For a given set of parameters entering $\bar{\Gamma}(\boldsymbol{k},\boldsymbol{k}')$, the solution of Eq.~\eqref{eq:eigenvalues_eq} with the highest eigenvalue $\lambda_{\rm max}$ will correspond to the highest critical temperature $T_c$. The symmetry of the pairing gap is reflected in the corresponding eigenfunction $\Delta_{\rm max}(\boldsymbol{k})$.

In the following, we consider the reference value of $U \approx 1.16$ eV, which has been estimated as $T_c \approx 80$ K  from real-space mean-field calculations   \cite{liuWavePairingDestructive2023}. Considering the largest interlayer hopping in the model ($|t_{\perp}^z| \approx 0.635$ eV) \cite{luoBilayerTwoOrbitalModel2023}, gives a moderate $U/|t_{\perp}^z|\approx 1.83$, reflecting an itinerant electron picture.
For these parameters and $U_\perp\!=\!0$, the leading and subleading solutions of Eq.~\eqref{eq:eigenvalues_eq}  display $s^{\pm}$-wave and $d_{xy}$-wave symmetries (with $\lambda^{s^{\pm}}/\lambda^{d_{xy}} > 1$), in accordance with previous studies \cite{yangPossiblewaveSuperconductivity2023,quBilayerModel2024,zhangStructuralPhaseTransition2024}.

For increasing values of $U_\perp$, however, the ratio $\lambda^{s^{\pm}}/\lambda^{d_{xy}}$ \emph{decreases} and eventually the solution with $d_{xy}$-wave symmetry becomes dominant. 
This is illustrated in Figure \ref{fig:phase_diagram}(a), which shows a color map of $\log{(\lambda^{s^{\pm}}/\lambda^{d_{xy}})}$ for different values of onsite $U$ and interlayer $U_\perp$ interaction strengths. 
The dashed line represents the curve $\lambda^{s^{\pm}}=\lambda^{d_{xy}}$ and can be interpreted as a ``phase boundary" separating the regions where each gap symmetry dominates.
The corresponding superconducting gap symmetry solutions projected onto the Fermi surface for the dominant $s^\pm$ and $d_{x^2-y^2}$ waves are shown in Fig.~\ref{fig:phase_diagram}(b) and (c), respectively.

\begin{figure}[t]
\begin{center}
\includegraphics[width=1\linewidth]{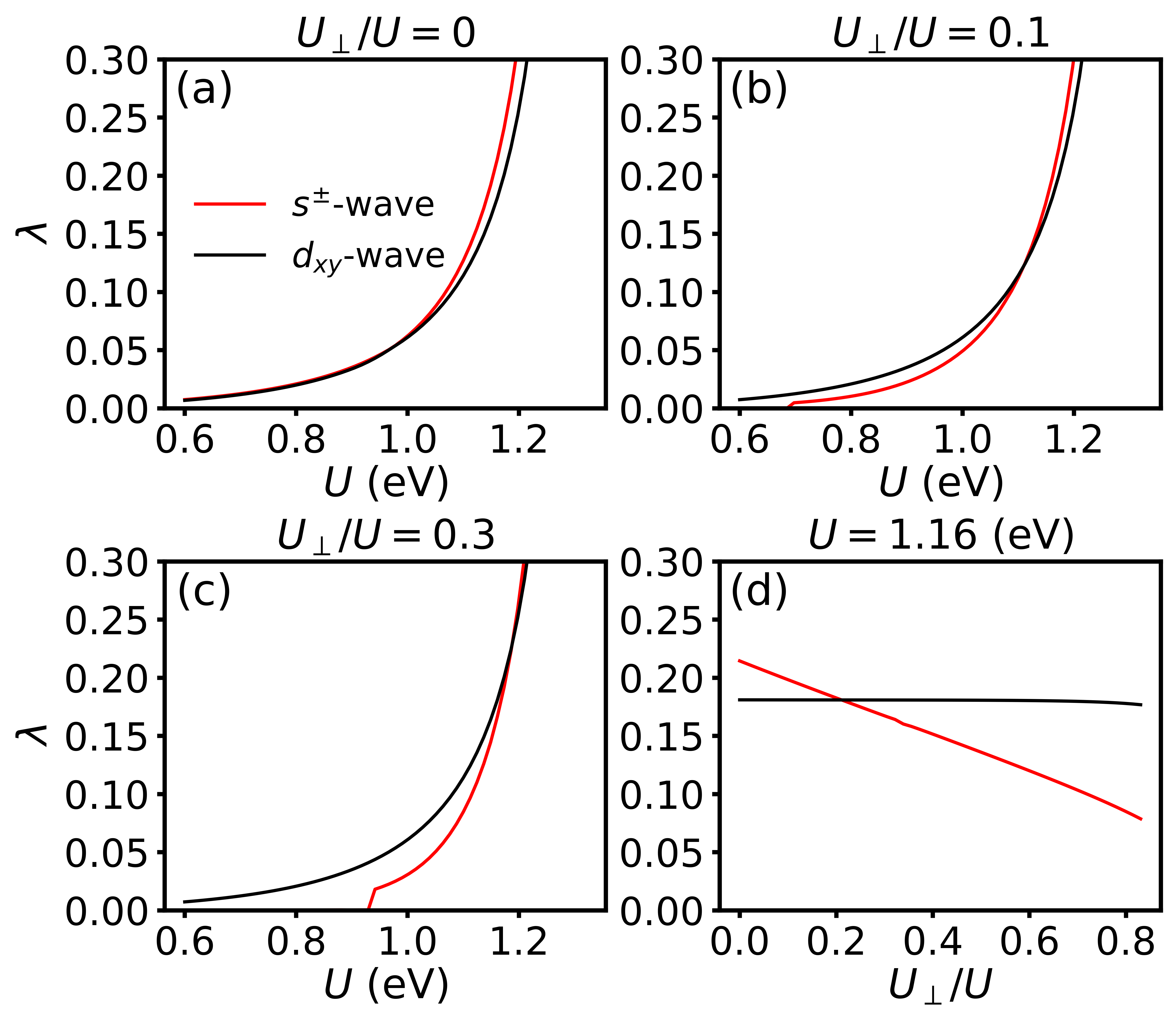}
\end{center}
\vspace{-17pt}
\caption{ Line-cuts of the phase diagram shown in Fig.~\ref{fig:phase_diagram} showing the leading and subleading eigenvalues of Eq.~\eqref{eq:eigenvalues_eq} as a function of (a,b,c) $U$ for different values of $U_{\perp}/U$ and (d)  $U_{\perp}/U$ for $U=1.16$ eV.  }
\label{fig:SI-cuts}
\end{figure}

A more detailed view of the leading and sub-leading solutions as a function of $U$ and $U_{\perp}$ is shown in Fig.~\ref{fig:SI-cuts}. For $U/U_\perp=0$ [Fig.~\ref{fig:SI-cuts}(a)], $s^\pm$ and $d_{x^2-y^2}$ waves are nearly degenerate for $U<1$ eV, while the $s^\pm$ dominates for larger values of $U$. By contrast,  increasing the ratio $U_\perp/U>0$ immediately breaks the degeneracy and suppresses the $s^\pm$ solution, although higher $U$ again favor the sign-changing $s$-wave, as shown in Figs.~\ref{fig:SI-cuts}(b) and (c).

For the reference value of $U \!=\! 1.16$ eV, we show in Fig.~\ref{fig:SI-cuts}(d) that the transition point occurs at a moderate value of $U^{c}_\perp \approx 0.2U = 0.23$ eV, with the gap functions displaying  $s^{\pm}$-wave and $d_{xy}$-wave symmetries for $U_\perp<U^{c}_\perp$ and $U_\perp>U^{c}_\perp$ [Figs. \ref{fig:phase_diagram}(b) and (c)], respectively. For larger values of $U$, however, the $s^{\pm}$-wave symmetry is clearly favored even for ratios $U_\perp/U$ as large as $50-60 \%$.

Additionally, as discussed in Appendix \ref{app:J_Jperp}, we note that the Hund's couplings $J$ and $J_{\perp}$ only marginally affect this general behavior. In fact, increasing $J/U$ does not influence the leading pairing symmetry [Fig.~\ref{fig:g_Jperp}]. Moreover, we find that changing $J_{\perp}/U_{\perp}$ does not change our conclusions qualitatively, although decreasing $J_{\perp}/U_{\perp}$ shifts up the phase boundary separating the two symmetries [dashed black line in Fig.~\ref{fig:phase_diagram}(a)].

\subsection{The role of the charge susceptibility}
\label{sec:chargesuscep}

A key aspect in the transition of $s^{\pm}$ to $d_{xy}$ pairing symmetries as a function of $U_{\perp}$ is its connection  the to the decrease in the charge susceptibility peaks near the nesting vectors, previously shown in Fig.~\ref{fig:Chi}.  As previously mentioned, Eq.~\eqref{eq:U_matrices} shows that the spin susceptibility is unaffected by the interlayer interactions. On the other hand, Figs.~\ref{fig:Chi}(c,d) show that the charge susceptibility is suppressed for increasing $U_\perp/U$ in regions around the $\boldsymbol{\Gamma}$ and $\boldsymbol{M}$ pockets. 

\begin{figure}[t]
\begin{center}
\includegraphics[width=1\linewidth]{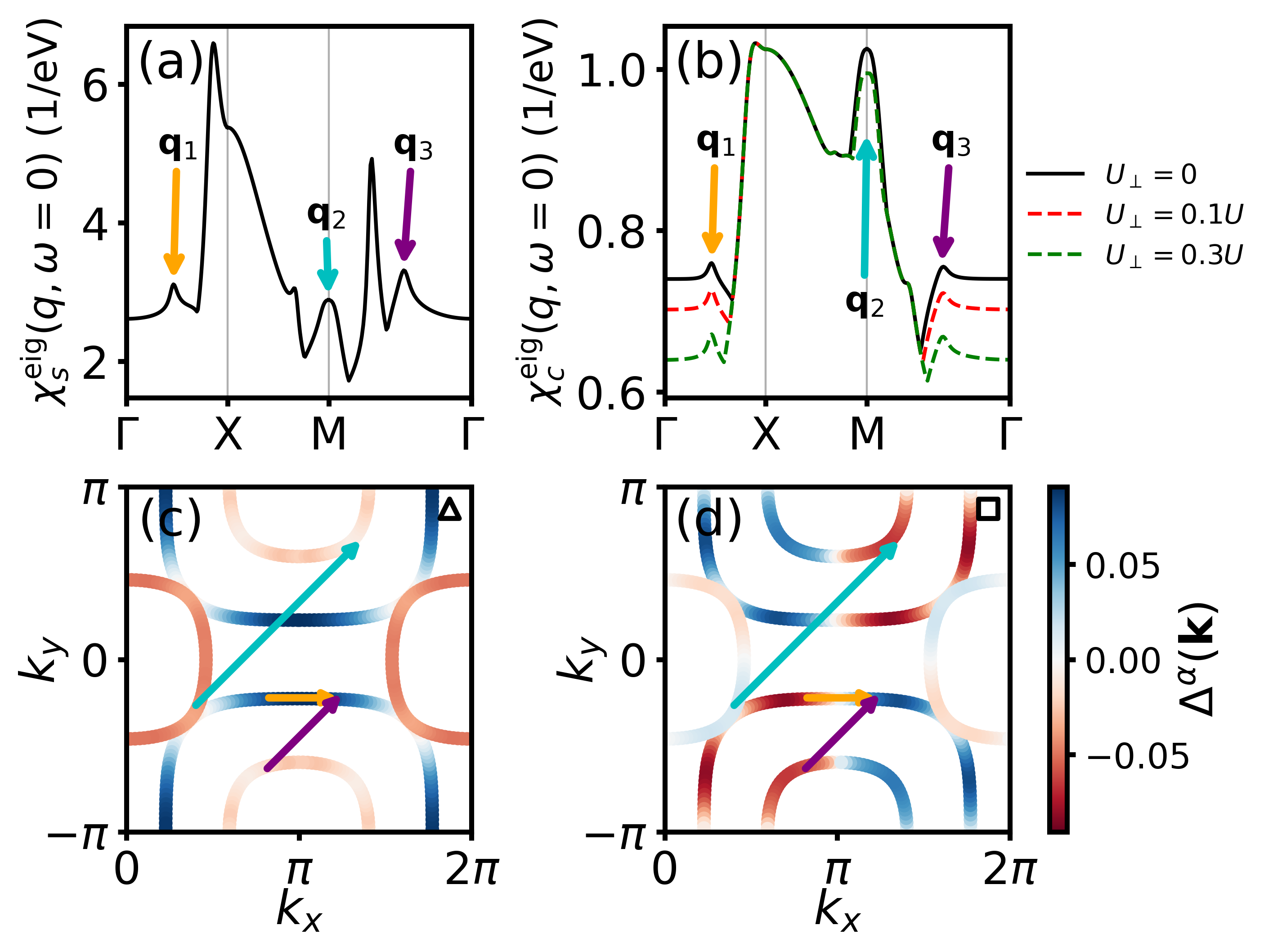}
\end{center}
\vspace{-17pt}
\caption{ Susceptibilities and the sign changes in the superconducting gap. The main eigenvalue of the spin and charge susceptibility matrices are shown in panels (a) and (b), respectively. Dashed lines represent results for different $U_{\perp}/U$ ratios, while the arrows point out to the susceptibility peaks related to each putative nesting vector $\boldsymbol{q}_i$. Panels (c) and (d) show the same gap function data of Fig.~\ref{fig:model_system}(b) and (c), respectively, while the colored arrows corresponded to the vectors $\boldsymbol{q}_i$ in the upper panels.
}
\label{fig:nesting}
\end{figure}

This is better illustrated in Figs.~\ref{fig:nesting}(a) and (b), which show the main eigenvalue of the spin [$\hat{\chi}_s(\boldsymbol{q},0)$] and charge [$\hat{\chi}_c(\boldsymbol{q},0)$] susceptibility matrices respectively along the $\boldsymbol{\Gamma}-\boldsymbol{X}-\boldsymbol{M}-\boldsymbol{\Gamma}$ high-symmetry path. The main three nesting vectors ($\boldsymbol{q}_1$, $\boldsymbol{q}_2$ and $\boldsymbol{q}_3$)  are marked  by small colored arrows.

Figures~\ref{fig:nesting}(c,d) show the same gap function data as Fig.~\ref{fig:model_system}(b,c) [$U_\perp/U=0.1$ ($\triangle$) and $U_\perp/U=0.3$ ($\square$), respectively]  with a shift in the $k_x$ component so that the three nesting vectors $\boldsymbol{q}_{i=1,2,3}$ (represented as colored arrows, in scale) can be better visualized. A comparison between Figs.~\ref{fig:nesting}(c) and (d) shows that the $\boldsymbol{q}_1$ and $\boldsymbol{q}_2$ nesting vectors connect pockets where the leading gap function has the \emph{same sign} in the $s^{\pm}$-wave case (i.e., $\Delta^{\alpha}(\boldsymbol{k})\!=\!+\Delta^{\alpha}(\boldsymbol{k}+\boldsymbol{q}_{i=1,2})$) while they are of \emph{ opposite sign} in the $d_{xy}$-wave case ($\Delta^{\alpha}(\boldsymbol{k})\!=\!-\Delta^{\alpha}(\boldsymbol{k}+\boldsymbol{q}_{i=1,2})$).

These results can be better understood by looking at both the dependence of the pairing vertex function on the charge susceptibility [Eq.~\eqref{eq:pairing_vertex}] and on the pairing function $\bar{\Gamma}(\boldsymbol{k},\boldsymbol{k}')$ on the left-hand-side of the linear pairing  equation [Eq.~\eqref{eq:eigenvalues_eq}]. First, we remark that $\hat{\chi}_s(\boldsymbol{q},0)$  contributes with a positive sign to the right-hand-side of Eq.~\eqref{eq:pairing_vertex}, while $\hat{\chi}_c(\boldsymbol{q},0)$ contributes with a negative sign.
As such, \emph{reductions} in $\hat{\chi}_c(\boldsymbol{q},0)$ tend to \emph{increase} the pairing $\bar{\Gamma}(\boldsymbol{k},\boldsymbol{k}+\boldsymbol{q})$, particularly at the nesting points $\boldsymbol{q}=\boldsymbol{q}_{1,2}$, at which the spin susceptibility is lower than at $\boldsymbol{q}=\boldsymbol{q}_{3}$ (see Fig.~\ref{fig:nesting}(a)).

Due to negative sign in the left-hand side in Eq.~\eqref{eq:eigenvalues_eq} and considering that $\bar{\Gamma}(\boldsymbol{k},\boldsymbol{k}+\boldsymbol{q})$ is positive at the nesting points, we argue that such increases in $\bar{\Gamma}(\boldsymbol{k},\boldsymbol{k}+\boldsymbol{q}_{1,2})$ as $U_{\perp}/U$ increases will favor sign-changing solutions which obey $\Delta^{\alpha}(\boldsymbol{k})\!=\!-\Delta^{\alpha}(\boldsymbol{k}+\boldsymbol{q}_{1,2})$ such as the $d_{xy}$-wave symmetric one.

\subsection{Orbital contributions to the pairing interaction}
\label{sec:orbitalcontributions}

Further insight as to why the interlayer interaction tends to favor a $d$-wave symmetry over the $s^\pm$ one can be gained by looking more closely on the role of the interorbital contribution to the pairing interaction.
Imagine if one could tune the relative contributions of the different FS sheets to the paring gap function defined by Eq.~\eqref{eq:eigenvalues_eq}. In practice, this can be accomplished by artificially giving different weights to the terms in the sum on the left-hand-side of Eq.~\eqref{eq:eigenvalues_eq}. For instance, one could exclude (assign zero weight to) the contribution of one of the FS pockets ($\alpha$, $\beta$, or $\gamma$) and see what the resulting symmetry of the leading pairing gap function looks like.

\begin{figure}[t]
\begin{center}
\includegraphics[width=0.98\linewidth]{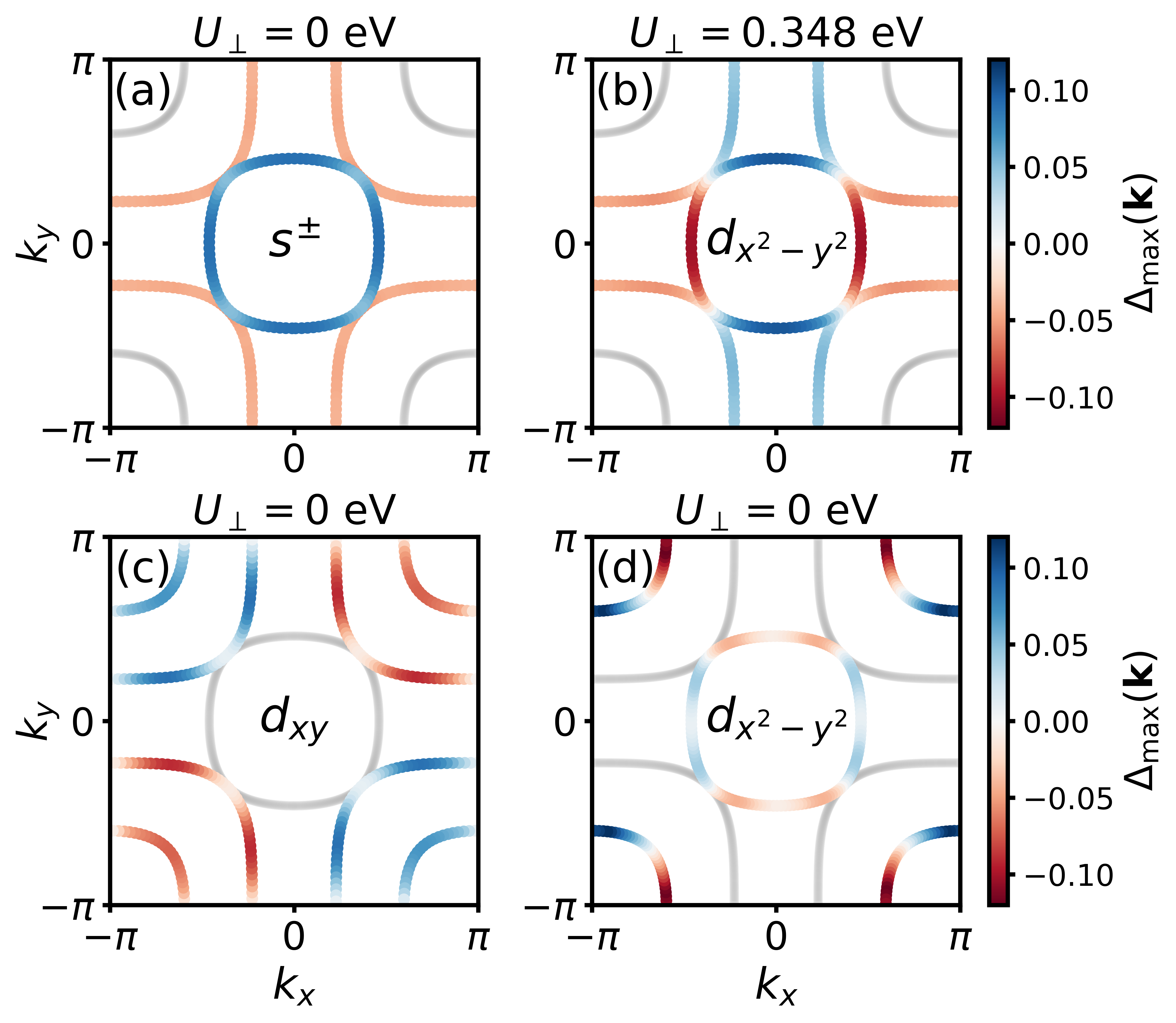}
\end{center}
\vspace{-17pt}
\caption{Color maps of the leading superconducting gap function $\Delta_{\rm max}(\boldsymbol{k})$ computed using the two-orbital bilayer pairing vertex but considering only the contributions of two out of the three FS pockets. Panels (a) and (b) shows results excluding the contributions of the $\gamma$ pocket for $U_\perp=0$ and $U_\perp=0.3 U=0.348$ eV, respectively. Panels (c) and (d) show results for $U_\perp=0$ excluding the contributions of $\alpha$ and $\beta$ pockets, respectively. In all cases, $U=1.16$ eV.
}
\label{fig:pockets}
\end{figure}

\begin{table}[t]
\caption{Summary of superconducting pairing symmetries found for each pocket or combination of pockets $\alpha,\beta,\gamma$ as indicated in Fig.~\ref{fig:model_system}(b). We set $U=1.16$ eV.}
\begin{ruledtabular}
\begin{tabular}{@{}ccc@{}}
Pockets         & $U_\perp/U$ & Symmetry           \\ \midrule
$\alpha$        & any                       & $g$-wave           \\
$\beta$         & any                       & $d_{xy}$-wave      \\
$\gamma$        & any                       & $d_{x^2-y^2}$-wave \\
$\alpha+\beta$  & $\lesssim0.3$             & $s^\pm$-wave       \\
$\alpha+\beta$  & $\gtrsim0.3$              & $d_{xy}$-wave      \\
$\alpha+\gamma$ & any                       & $d_{x^2-y^2}$-wave \\
$\beta+\gamma$  & any                       & $d_{xy}$-wave      \\
all             & $\lesssim0.2$             & $s^\pm$-wave       \\
all             & $\gtrsim0.2$              & $d_{xy}$-wave      \\ 
\end{tabular}
\end{ruledtabular}
\label{tab:symmetries}
\end{table}

This is done in Fig.~\ref{fig:pockets} for $U=1.16$ eV. Figure \ref{fig:pockets}(a) shows that, for $U_\perp=0$, the paring gap symmetry remains $s^{\pm}$-wave even if the contribution of the $\gamma$ pocket is excluded and only pockets $\alpha$ and $\beta$, which are dominated by $3 d_{x^2-y^2}$ orbitals and hence have a more ``planar'' character, are considered in the calculation. In a sense, this would mimic a ``cuprate-like'' scenario  \cite{fanSuperconductivityNickelateCuprate2024}. In this situation, by increasing $U_\perp$ the symmetry of the leading gap function becomes $d_{x^2-y^2}$ and \emph{not} $d_{xy}$, as in the case where all pockets are considered [Fig.~\ref{fig:phase_diagram}(c)]. On the other hand, the pairing symmetry becomes $d_{xy}$ if, instead, the contribution of the $\alpha$  pocket is excluded [Fig.~\ref{fig:pockets}(c)]. Finally, if the contribution of the $\beta$ pocket is excluded, the gap symmetry becomes $d_{x^2-y^2}$ [Fig.~\ref{fig:pockets}(d)] and not $d_{xy}$. A summary of these findings  is presented in Table~\ref{tab:symmetries}, which also includes other combinations of the $\alpha,\beta,\gamma$ pockets  (as detailed in Appendix \ref{app:pockets}).

These results indicate that, when all three pockets are considered in the calculation (Fig.~\ref{fig:phase_diagram}), by increasing  $U_\perp$, the relative contribution of the $\gamma$ pocket (dominated by $d_{3z^2-r^2}$ orbitals) increases over the (more planar) $\alpha$ pocket contribution, thereby inducing the switching between $s^{\pm}$- and nodal $d_{xy}$-wave paring symmetries. As such, the interplay between $\gamma$ and $\alpha$ pockets in the pairing vertex is a crucial element in explaining the $s^\pm \rightarrow d_{xy}$ order switching as $U_\perp$ increases.  

\begin{figure}[t]
\begin{center}
\includegraphics[width=1\linewidth]{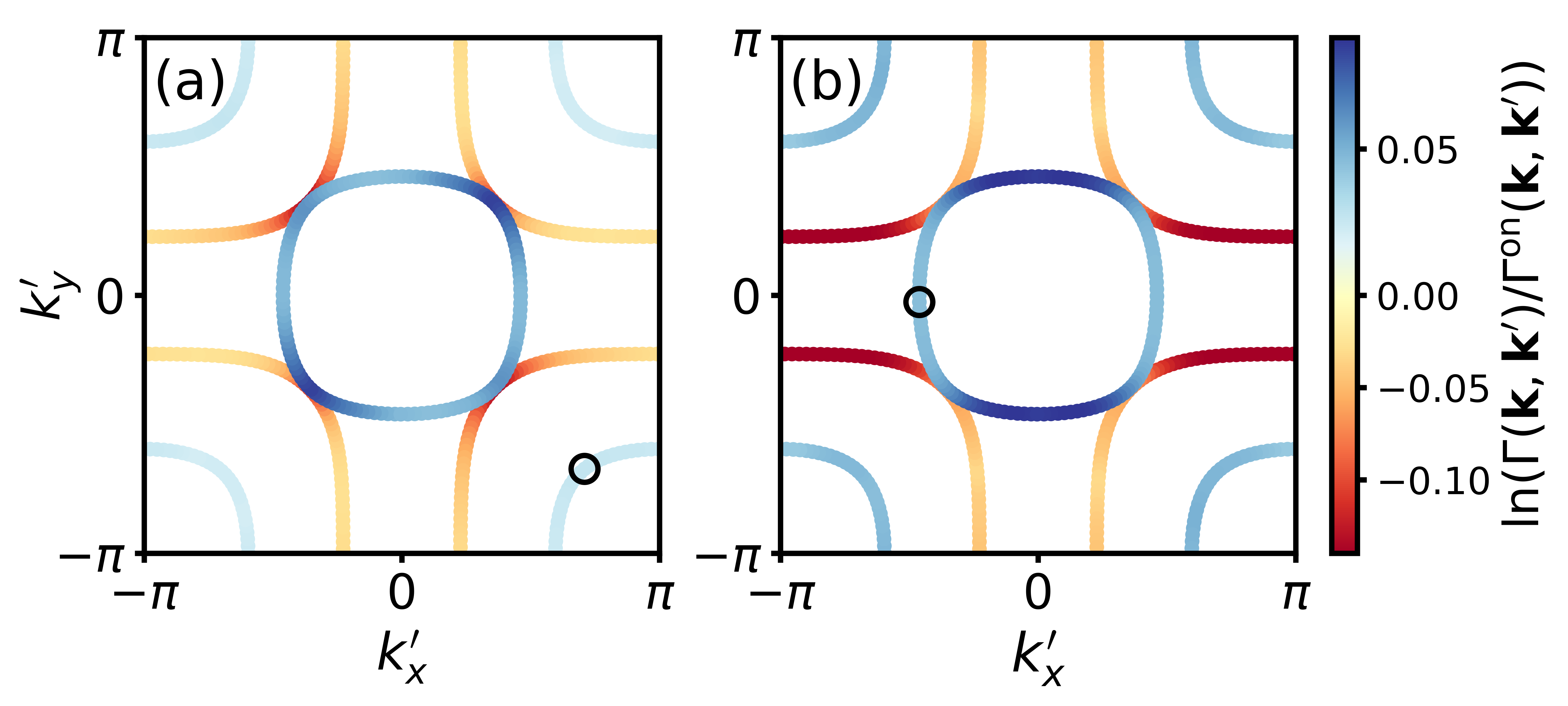}
\end{center}
\vspace{-17pt}
\caption{ Enhancement of the pairing vertex function ($\Gamma(\boldsymbol{k},\boldsymbol{k}')$) by interlayer interactions. Panels (a) and (b) show the ratio of $\Gamma(\boldsymbol{k},\boldsymbol{k}')$ calculated with $U_{\perp}=0.348$ eV  and $\Gamma^{\mathrm{on}}(\boldsymbol{k},\boldsymbol{k}')$, calculated with $U_{\perp}=0$  in a logarithmic scale for a fixed $\boldsymbol{k}$ (black ring) 
located at (a) $\gamma$ and (b) $\alpha$ FS pockets, dominated by $d_{3z^2-r^2}$ and $d_{x^2-y^2}$ orbitals respectively.
}
\label{fig:pairing_int}
\end{figure}

This conclusion is further supported by looking at the differences in the pairing vertex function $\Gamma(\boldsymbol{k},\boldsymbol{k}')$ for $U_{\perp}=0.348$ eV and $U_{\perp}=0$. Figure~\ref{fig:pairing_int} shows 
the ratio between $\Gamma(\boldsymbol{k},\boldsymbol{k}')$ (calculated with $U_{\perp}=0.348$ eV) and $\Gamma^{\rm on}(\boldsymbol{k},\boldsymbol{k}')$ (calculated with $U_{\perp}=0$), for a fixed $\boldsymbol{k}$ (marked as a black circle) and as a function of $\boldsymbol{k}'$ along the FS. 

For example, as shown in Fig.~\ref{fig:pairing_int}(a), for a $\boldsymbol{k}$-point located in the $\gamma$ pocket, increasing $U_{\perp}$ leads to an \emph{enhancement} of the pairing vertex function [$\Gamma(\boldsymbol{k},\boldsymbol{k}') > \Gamma^{\rm on}(\boldsymbol{k},\boldsymbol{k}')$] along the $\alpha$ pocket and a \emph{decrease} [$\Gamma(\boldsymbol{k},\boldsymbol{k}') < \Gamma^{\rm on}(\boldsymbol{k},\boldsymbol{k}')$] along the $\beta$ pocket. A similar result occurs when $\boldsymbol{k}$ is located at the $\alpha$ pocket [Fig,~\ref{fig:pairing_int}(b)]: the pairing vertex function is enhanced along the $\gamma$ pocket and decreases along the $\beta$ pocket. 

This result is consistent with our picture that interlayer interactions tend to favor the $d_{xy}$ pairing symmetry over the $s^\pm$ by favoring the interorbital pairing between electrons in the $\gamma$ and $\alpha$ pockets, which are dominated by the $d_{3z^2-r^2}$ and $d_{x^2-y^2}$ orbital contributions, respectively.

\section{Concluding remarks}
\label{sec:Conclusions}
In this work, we investigate the role of interlayer Coulomb interaction terms on
the superconducting properties of $\text{La}_3\text{Ni}_2\text{O}_7$ under pressure in the weak- to intermediate-coupling limit. 
Our results indicate that for moderate onsite interactions in the model (up to $U \sim 1.1$ eV), and relatively small interlayer interaction strengths (about $\sim 30\%$ of the onsite term) the leading pairing symmetry becomes of the $d_{xy}$-wave type instead of the $s^{\pm}$-wave predicted by several works in the \emph{absence} of interlayer interactions 
\cite{zhangStructuralPhaseTransition2024,liuWavePairingDestructive2023,zhangStructuralPhaseTransition2024,quBilayerModel2024,
chenOrbitalselectiveSuperconductivityPressurized2024}.
Our calculations also confirm that the $s^{\pm}$ solution is dominant for large onsite interactions ($U \gtrsim 1.3$ eV). This is consistent with results from strong-coupling ($t\!-\!\tilde{J}\!-\!\tilde{J}_{\perp}$) bi-layer single-orbital  models which show $s^\pm$-wave superconductivity favored by a strong interlayer exchange coupling \cite{quBilayerModel2024, chenOrbitalselectiveSuperconductivityPressurized2024}.

Additionally, our results indicate that charge fluctuations play an essential role in the $s$- to $-d$-wave transition as the interlayer interaction $U_{\perp}$ increases. This is a consequence of a decrease in the charge susceptibility  for large $U_{\perp}$ values, particularly at momenta $\boldsymbol{q}_{i}$ satisfying the nesting condition between different FS sheets. This in turn leads to an  \emph{increase} of the RPA paring function at these nesting vectors, favoring $d_{xy}$-wave sign-changing solutions obeying $\Delta^{\alpha}(\boldsymbol{k})\!=\!-\Delta^{\alpha}(\boldsymbol{k}+\boldsymbol{q}_{i})$ over the $s^{\pm}$-wave ones. We note that these $s^{\pm}$-wave solutions differ from those appearing in iron-based superconductors \cite{graserDegeneracySeveralPairing2009,kemperSensitivitySuperconductingState2010}, for which sign changes can be traced back to the \emph{spin} susceptibility at the nesting points. 

\begin{figure}[t]
\begin{center}
\includegraphics[width=1\linewidth]{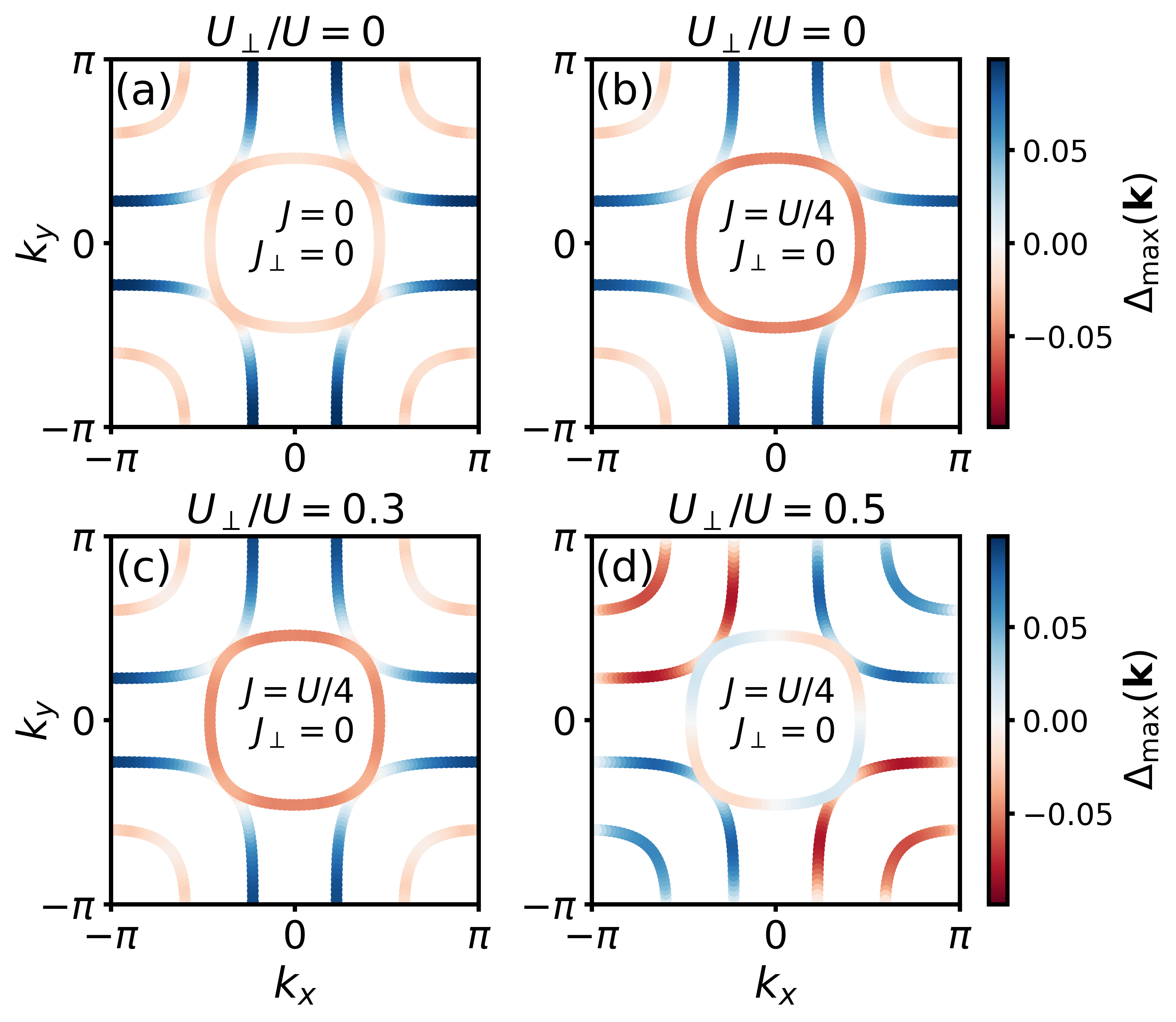}
\end{center}
\vspace{-17pt}
\caption{ Leading gap functions $\Delta_{\rm max}(\boldsymbol{k})$ displaying $s^{\pm}$-wave and $d_{xy}$-wave symmetries for $U=1.16$ eV and $U_\perp/U=0.1$ ($\triangle$) and $U_\perp/U=0.3$ ($\square$).  }
\label{fig:g_Jperp}
\end{figure}

More interestingly, our analysis shows that the superconducting pairing symmetry is strongly dependent on the interplay of \emph{all three} FS pockets. In particular, the interlayer interaction tends to favor interorbital pairing between electrons with $d_{3z^2-r^2}$ and $d_{x^2-y^2}$ orbital character, underscoring the view that a complete description of the superconducting properties can only be achieved by multi-orbital models \cite{zhangStructuralPhaseTransition2024,luoHighTCSuperconductivityLa3Ni2O72024,fanSuperconductivityNickelateCuprate2024,lechermannElectronicCorrelationsSuperconducting2023}.
In fact, we expect that a similar picture can emerge in systems exhibiting multiple FS pockets with close competing pairing symmetries such as the three-layer La$_4$Ni$_3$O$_{10}$ \cite{Zhang:Phys.Rev.Lett.:136001:2024}.

\emph{Note added:} After the submission of this manuscript, we became aware of  related works \cite{Xi_Phys.Rev.B_104505_2025,Zhan:arXiv:2503.18877:2025} 
reporting transitions from $s$-wave to $d$-wave pairing symmetry as a function of the interlayer interaction strength. Although these works use different methods (namely FLEX \cite{Xi_Phys.Rev.B_104505_2025} and fRG \cite{Zhan:arXiv:2503.18877:2025}), the main results are qualitatively similar to our findings.

\section{Acknowledgments} 
We thank Ilya Eremin and Steffen B\"otzel for fruitful discussions.
We acknowledge financial support from the S\~ao Paulo Research Foundation (FAPESP), Brazil (2022/15453-0 and 2023/14902-8) and from CNPq (Grant No. 312622/2023-6).

\section{Data availability}
The data that support the findings of this paper are openly available on Zenodo \cite{ZenodoData}.

\appendix

\begin{figure}[t]
\begin{center}
\includegraphics[width=1\linewidth]{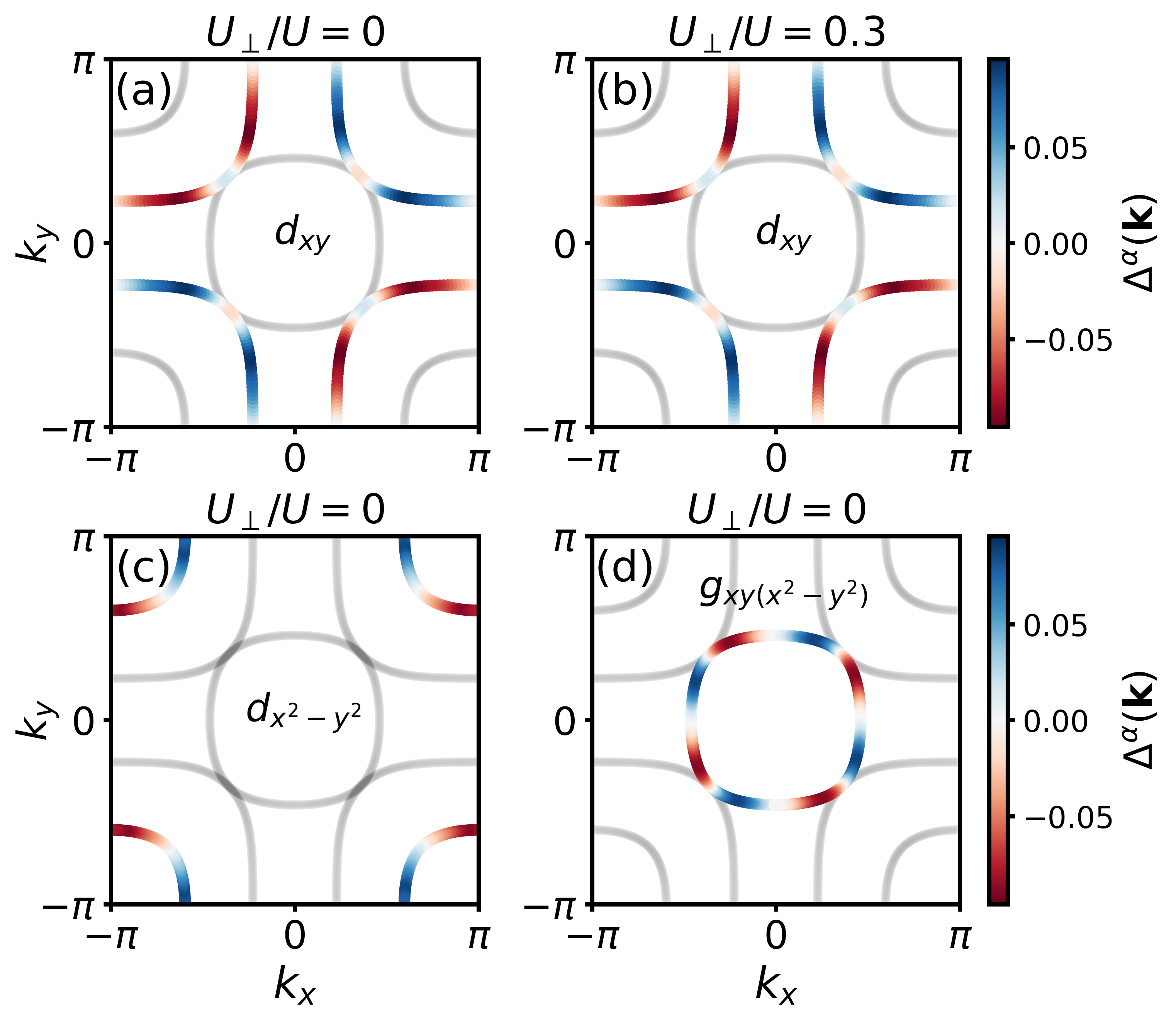}
\end{center}
\vspace{-17pt}
\caption{ Same as Fig.~\ref{fig:pockets} but for individual pockets $\beta$ (a,b), $\gamma$ (c), and $\alpha$ (d). }
\label{fig:SI-pockets}
\end{figure}

\section{Influence of Hund's couplings in the pairing symmetry}
\label{app:J_Jperp}
The influence of Hund's couplings $J$ and $J_\perp$ in the pairing symmetry is shown in Fig.~\ref{fig:g_Jperp}.
For $U/U_\perp=0$, Fig.~\ref{fig:g_Jperp}(a) shows that $J=J_\perp=0$ results in a leading $s^\pm$-wave, while increasing $J/U$ does not influence the pairing symmetry but for enhancing the gap function at the $\alpha$ pocket in comparison with the other pockets, which is illustrated in panel (b).
On the other hand, fixing $J=U/4$ and $J_\perp=0$, the $U_\perp/U=0.3$ gap symmetry shown in Fig.~\ref{fig:g_Jperp}(c) results in a leading $s^\pm$-wave, while only further increase of interlayer interactions, $U_\perp/U=0.5$ as in Fig.~\ref{fig:g_Jperp}(d), induces the subleading $d_{xy}$ as the leading pairing symmetry.
Therefore, lowering $J_\perp/U_\perp$ only increases the ratio $U_\perp/U$ necessary for the $s^\pm\rightarrow d_{xy}$ transition, which does not qualitatively change the phase diagram of Fig.~\ref{fig:phase_diagram}(a).

\section{Pairing symmetry of individual pockets}
\label{app:pockets}
Fig.~\ref{fig:pockets} shows results for couples of pockets but we have also checked that conclusions hold for the individual pockets as well.
In Fig.~\ref{fig:SI-pockets}, panels (a) and (b) show no change in the $d_{xy}$ symmetry of the $\beta$ pocket as the interlayer coupling varies from $U_\perp/U=0$ (a) to $U_\perp/U=0.3$ (b).
In fact, the interlayer interactions do not change the symmetry in any three cases explored.
As shown in panel (c), a $d_{x^2-y^2}$-wave dominates pairing of individual $\gamma$ pocket, and a $g$-wave symmetry is obtained for the $\alpha$ pocket alone (d).
None of the single-pocket solutions reproduces the $s^\pm\rightarrow d_{xy}$ transition as a function of interlayer interactions.


\begin{thebibliography}{37}%
\makeatletter
\providecommand \@ifxundefined [1]{%
 \@ifx{#1\undefined}
}%
\providecommand \@ifnum [1]{%
 \ifnum #1\expandafter \@firstoftwo
 \else \expandafter \@secondoftwo
 \fi
}%
\providecommand \@ifx [1]{%
 \ifx #1\expandafter \@firstoftwo
 \else \expandafter \@secondoftwo
 \fi
}%
\providecommand \natexlab [1]{#1}%
\providecommand \enquote  [1]{``#1''}%
\providecommand \bibnamefont  [1]{#1}%
\providecommand \bibfnamefont [1]{#1}%
\providecommand \citenamefont [1]{#1}%
\providecommand \href@noop [0]{\@secondoftwo}%
\providecommand \href [0]{\begingroup \@sanitize@url \@href}%
\providecommand \@href[1]{\@@startlink{#1}\@@href}%
\providecommand \@@href[1]{\endgroup#1\@@endlink}%
\providecommand \@sanitize@url [0]{\catcode `\\12\catcode `\$12\catcode
  `\&12\catcode `\#12\catcode `\^12\catcode `\_12\catcode `\%12\relax}%
\providecommand \@@startlink[1]{}%
\providecommand \@@endlink[0]{}%
\providecommand \url  [0]{\begingroup\@sanitize@url \@url }%
\providecommand \@url [1]{\endgroup\@href {#1}{\urlprefix }}%
\providecommand \urlprefix  [0]{URL }%
\providecommand \Eprint [0]{\href }%
\providecommand \doibase [0]{https://doi.org/}%
\providecommand \selectlanguage [0]{\@gobble}%
\providecommand \bibinfo  [0]{\@secondoftwo}%
\providecommand \bibfield  [0]{\@secondoftwo}%
\providecommand \translation [1]{[#1]}%
\providecommand \BibitemOpen [0]{}%
\providecommand \bibitemStop [0]{}%
\providecommand \bibitemNoStop [0]{.\EOS\space}%
\providecommand \EOS [0]{\spacefactor3000\relax}%
\providecommand \BibitemShut  [1]{\csname bibitem#1\endcsname}%
\let\auto@bib@innerbib\@empty
\bibitem [{\citenamefont {Sun}\ \emph {et~al.}(2023)\citenamefont {Sun},
  \citenamefont {Huo}, \citenamefont {Hu}, \citenamefont {Li}, \citenamefont
  {Liu}, \citenamefont {Han}, \citenamefont {Tang}, \citenamefont {Mao},
  \citenamefont {Yang}, \citenamefont {Wang}, \citenamefont {Cheng},
  \citenamefont {Yao}, \citenamefont {Zhang},\ and\ \citenamefont
  {Wang}}]{sunSignaturesSuperconductivity802023}%
  \BibitemOpen
  \bibfield  {author} {\bibinfo {author} {\bibfnamefont {H.}~\bibnamefont
  {Sun}}, \bibinfo {author} {\bibfnamefont {M.}~\bibnamefont {Huo}}, \bibinfo
  {author} {\bibfnamefont {X.}~\bibnamefont {Hu}}, \bibinfo {author}
  {\bibfnamefont {J.}~\bibnamefont {Li}}, \bibinfo {author} {\bibfnamefont
  {Z.}~\bibnamefont {Liu}}, \bibinfo {author} {\bibfnamefont {Y.}~\bibnamefont
  {Han}}, \bibinfo {author} {\bibfnamefont {L.}~\bibnamefont {Tang}}, \bibinfo
  {author} {\bibfnamefont {Z.}~\bibnamefont {Mao}}, \bibinfo {author}
  {\bibfnamefont {P.}~\bibnamefont {Yang}}, \bibinfo {author} {\bibfnamefont
  {B.}~\bibnamefont {Wang}}, \bibinfo {author} {\bibfnamefont {J.}~\bibnamefont
  {Cheng}}, \bibinfo {author} {\bibfnamefont {D.-X.}\ \bibnamefont {Yao}},
  \bibinfo {author} {\bibfnamefont {G.-M.}\ \bibnamefont {Zhang}},\ and\
  \bibinfo {author} {\bibfnamefont {M.}~\bibnamefont {Wang}},\ }\bibfield
  {title} {\bibinfo {title} {Signatures of superconductivity near 80 {{K}} in a
  nickelate under high pressure},\ }\href
  {https://doi.org/10.1038/s41586-023-06408-7} {\bibfield  {journal} {\bibinfo
  {journal} {Nature}\ }\textbf {\bibinfo {volume} {621}},\ \bibinfo {pages}
  {493} (\bibinfo {year} {2023})}\BibitemShut {NoStop}%
\bibitem [{\citenamefont {Wang}\ \emph
  {et~al.}(2024{\natexlab{a}})\citenamefont {Wang}, \citenamefont {Wang},
  \citenamefont {Shen}, \citenamefont {Hou}, \citenamefont {Ma}, \citenamefont
  {Shi}, \citenamefont {Ren}, \citenamefont {Gu}, \citenamefont {Ma},
  \citenamefont {Yang}, \citenamefont {Liu}, \citenamefont {Guo}, \citenamefont
  {Sun}, \citenamefont {Zhang}, \citenamefont {Calder}, \citenamefont {Yan},
  \citenamefont {Wang}, \citenamefont {Uwatoko},\ and\ \citenamefont
  {Cheng}}]{wangPressureInducedSuperconductivityPolycrystalline2024}%
  \BibitemOpen
  \bibfield  {author} {\bibinfo {author} {\bibfnamefont {G.}~\bibnamefont
  {Wang}}, \bibinfo {author} {\bibfnamefont {N.~N.}\ \bibnamefont {Wang}},
  \bibinfo {author} {\bibfnamefont {X.~L.}\ \bibnamefont {Shen}}, \bibinfo
  {author} {\bibfnamefont {J.}~\bibnamefont {Hou}}, \bibinfo {author}
  {\bibfnamefont {L.}~\bibnamefont {Ma}}, \bibinfo {author} {\bibfnamefont
  {L.~F.}\ \bibnamefont {Shi}}, \bibinfo {author} {\bibfnamefont {Z.~A.}\
  \bibnamefont {Ren}}, \bibinfo {author} {\bibfnamefont {Y.~D.}\ \bibnamefont
  {Gu}}, \bibinfo {author} {\bibfnamefont {H.~M.}\ \bibnamefont {Ma}}, \bibinfo
  {author} {\bibfnamefont {P.~T.}\ \bibnamefont {Yang}}, \bibinfo {author}
  {\bibfnamefont {Z.~Y.}\ \bibnamefont {Liu}}, \bibinfo {author} {\bibfnamefont
  {H.~Z.}\ \bibnamefont {Guo}}, \bibinfo {author} {\bibfnamefont {J.~P.}\
  \bibnamefont {Sun}}, \bibinfo {author} {\bibfnamefont {G.~M.}\ \bibnamefont
  {Zhang}}, \bibinfo {author} {\bibfnamefont {S.}~\bibnamefont {Calder}},
  \bibinfo {author} {\bibfnamefont {J.-Q.}\ \bibnamefont {Yan}}, \bibinfo
  {author} {\bibfnamefont {B.~S.}\ \bibnamefont {Wang}}, \bibinfo {author}
  {\bibfnamefont {Y.}~\bibnamefont {Uwatoko}},\ and\ \bibinfo {author}
  {\bibfnamefont {J.-G.}\ \bibnamefont {Cheng}},\ }\bibfield  {title} {\bibinfo
  {title} {Pressure-induced superconductivity in polycrystalline
  ${\mathrm{la}}_{3}{\mathrm{ni}}_{2}{\mathrm{o}}_{7\ensuremath{-}\ensuremath{\delta}}$},\
  }\href {https://doi.org/10.1103/PhysRevX.14.011040} {\bibfield  {journal}
  {\bibinfo  {journal} {Physical Review X}\ }\textbf {\bibinfo {volume} {14}},\
  \bibinfo {pages} {011040} (\bibinfo {year} {2024}{\natexlab{a}})}\BibitemShut
  {NoStop}%
\bibitem [{\citenamefont {Li}\ \emph {et~al.}(2024{\natexlab{a}})\citenamefont
  {Li}, \citenamefont {Ma}, \citenamefont {Zhang}, \citenamefont {Huang},
  \citenamefont {Huang}, \citenamefont {Huo}, \citenamefont {Hu}, \citenamefont
  {Dong}, \citenamefont {He}, \citenamefont {Liao}, \citenamefont {Chen},
  \citenamefont {Xie}, \citenamefont {Sun},\ and\ \citenamefont
  {Wang}}]{Li:NickelateMeissner:arXiv:2024}%
  \BibitemOpen
  \bibfield  {author} {\bibinfo {author} {\bibfnamefont {J.}~\bibnamefont
  {Li}}, \bibinfo {author} {\bibfnamefont {P.}~\bibnamefont {Ma}}, \bibinfo
  {author} {\bibfnamefont {H.}~\bibnamefont {Zhang}}, \bibinfo {author}
  {\bibfnamefont {X.}~\bibnamefont {Huang}}, \bibinfo {author} {\bibfnamefont
  {C.}~\bibnamefont {Huang}}, \bibinfo {author} {\bibfnamefont
  {M.}~\bibnamefont {Huo}}, \bibinfo {author} {\bibfnamefont {D.}~\bibnamefont
  {Hu}}, \bibinfo {author} {\bibfnamefont {Z.}~\bibnamefont {Dong}}, \bibinfo
  {author} {\bibfnamefont {C.}~\bibnamefont {He}}, \bibinfo {author}
  {\bibfnamefont {J.}~\bibnamefont {Liao}}, \bibinfo {author} {\bibfnamefont
  {X.}~\bibnamefont {Chen}}, \bibinfo {author} {\bibfnamefont {T.}~\bibnamefont
  {Xie}}, \bibinfo {author} {\bibfnamefont {H.}~\bibnamefont {Sun}},\ and\
  \bibinfo {author} {\bibfnamefont {M.}~\bibnamefont {Wang}},\ }\bibfield
  {title} {\bibinfo {title} {Pressure-driven right-triangle shape
  superconductivity in bilayer nickelate la$_3$ni$_2$o$_7$},\ }\href
  {https://arxiv.org/abs/2404.11369} {\bibfield  {journal} {\bibinfo  {journal}
  {arXiv}\ ,\ \bibinfo {pages} {2404.11369}} (\bibinfo {year}
  {2024}{\natexlab{a}})}\BibitemShut {NoStop}%
\bibitem [{\citenamefont {Wang}\ \emph
  {et~al.}(2024{\natexlab{b}})\citenamefont {Wang}, \citenamefont {Wen},
  \citenamefont {Wu}, \citenamefont {Yao},\ and\ \citenamefont
  {Xiang}}]{WangReview:ChinesePhysicsLetters:077402:2024}%
  \BibitemOpen
  \bibfield  {author} {\bibinfo {author} {\bibfnamefont {M.}~\bibnamefont
  {Wang}}, \bibinfo {author} {\bibfnamefont {H.-H.}\ \bibnamefont {Wen}},
  \bibinfo {author} {\bibfnamefont {T.}~\bibnamefont {Wu}}, \bibinfo {author}
  {\bibfnamefont {D.-X.}\ \bibnamefont {Yao}},\ and\ \bibinfo {author}
  {\bibfnamefont {T.}~\bibnamefont {Xiang}},\ }\bibfield  {title} {\bibinfo
  {title} {Normal and superconducting properties of la3ni2o7},\ }\href
  {https://doi.org/10.1088/0256-307X/41/7/077402} {\bibfield  {journal}
  {\bibinfo  {journal} {Chinese Physics Letters}\ }\textbf {\bibinfo {volume}
  {41}},\ \bibinfo {pages} {077402} (\bibinfo {year}
  {2024}{\natexlab{b}})}\BibitemShut {NoStop}%
\bibitem [{\citenamefont {Ko}\ \emph {et~al.}(2024)\citenamefont {Ko},
  \citenamefont {Yu}, \citenamefont {Liu}, \citenamefont {Bhatt}, \citenamefont
  {Li}, \citenamefont {Thampy}, \citenamefont {Kuo}, \citenamefont {Wang},
  \citenamefont {Lee}, \citenamefont {Lee}, \citenamefont {Lee}, \citenamefont
  {Goodge}, \citenamefont {Muller},\ and\ \citenamefont
  {Hwang}}]{Ko:Nature::2024}%
  \BibitemOpen
  \bibfield  {author} {\bibinfo {author} {\bibfnamefont {E.~K.}\ \bibnamefont
  {Ko}}, \bibinfo {author} {\bibfnamefont {Y.}~\bibnamefont {Yu}}, \bibinfo
  {author} {\bibfnamefont {Y.}~\bibnamefont {Liu}}, \bibinfo {author}
  {\bibfnamefont {L.}~\bibnamefont {Bhatt}}, \bibinfo {author} {\bibfnamefont
  {J.}~\bibnamefont {Li}}, \bibinfo {author} {\bibfnamefont {V.}~\bibnamefont
  {Thampy}}, \bibinfo {author} {\bibfnamefont {C.-T.}\ \bibnamefont {Kuo}},
  \bibinfo {author} {\bibfnamefont {B.~Y.}\ \bibnamefont {Wang}}, \bibinfo
  {author} {\bibfnamefont {Y.}~\bibnamefont {Lee}}, \bibinfo {author}
  {\bibfnamefont {K.}~\bibnamefont {Lee}}, \bibinfo {author} {\bibfnamefont
  {J.-S.}\ \bibnamefont {Lee}}, \bibinfo {author} {\bibfnamefont {B.~H.}\
  \bibnamefont {Goodge}}, \bibinfo {author} {\bibfnamefont {D.~A.}\
  \bibnamefont {Muller}},\ and\ \bibinfo {author} {\bibfnamefont {H.~Y.}\
  \bibnamefont {Hwang}},\ }\bibfield  {title} {\bibinfo {title} {Signatures of
  ambient pressure superconductivity in thin film la3ni2o7},\ }\href
  {https://doi.org/10.1038/s41586-024-08525-3} {\bibfield  {journal} {\bibinfo
  {journal} {Nature}\ } (\bibinfo {year} {2024})}\BibitemShut {NoStop}%
\bibitem [{\citenamefont {Luo}\ \emph {et~al.}(2023)\citenamefont {Luo},
  \citenamefont {Hu}, \citenamefont {Wang}, \citenamefont {W{\'u}},\ and\
  \citenamefont {Yao}}]{luoBilayerTwoOrbitalModel2023}%
  \BibitemOpen
  \bibfield  {author} {\bibinfo {author} {\bibfnamefont {Z.}~\bibnamefont
  {Luo}}, \bibinfo {author} {\bibfnamefont {X.}~\bibnamefont {Hu}}, \bibinfo
  {author} {\bibfnamefont {M.}~\bibnamefont {Wang}}, \bibinfo {author}
  {\bibfnamefont {W.}~\bibnamefont {W{\'u}}},\ and\ \bibinfo {author}
  {\bibfnamefont {D.-X.}\ \bibnamefont {Yao}},\ }\bibfield  {title} {\bibinfo
  {title} {Bilayer two-orbital model of
  $\mathrm{L}{\mathrm{a}}_{3}\mathrm{N}{\mathrm{i}}_{2}{\mathrm{o}}_{7}$ under
  pressure},\ }\href {https://doi.org/10.1103/PhysRevLett.131.126001}
  {\bibfield  {journal} {\bibinfo  {journal} {Physical Review Letters}\
  }\textbf {\bibinfo {volume} {131}},\ \bibinfo {pages} {126001} (\bibinfo
  {year} {2023})}\BibitemShut {NoStop}%
\bibitem [{\citenamefont {Zhang}\ \emph {et~al.}(2023)\citenamefont {Zhang},
  \citenamefont {Lin}, \citenamefont {Moreo},\ and\ \citenamefont
  {Dagotto}}]{zhangElectronicStructureDimer2023}%
  \BibitemOpen
  \bibfield  {author} {\bibinfo {author} {\bibfnamefont {Y.}~\bibnamefont
  {Zhang}}, \bibinfo {author} {\bibfnamefont {L.-F.}\ \bibnamefont {Lin}},
  \bibinfo {author} {\bibfnamefont {A.}~\bibnamefont {Moreo}},\ and\ \bibinfo
  {author} {\bibfnamefont {E.}~\bibnamefont {Dagotto}},\ }\bibfield  {title}
  {\bibinfo {title} {Electronic structure, dimer physics, orbital-selective
  behavior, and magnetic tendencies in the bilayer nickelate superconductor
  ${\mathrm{la}}_{3}{\mathrm{ni}}_{2}{\mathrm{o}}_{7}$ under pressure},\ }\href
  {https://doi.org/10.1103/PhysRevB.108.L180510} {\bibfield  {journal}
  {\bibinfo  {journal} {Physical Review B}\ }\textbf {\bibinfo {volume}
  {108}},\ \bibinfo {pages} {L180510} (\bibinfo {year} {2023})}\BibitemShut
  {NoStop}%
\bibitem [{\citenamefont {Zhang}\ \emph
  {et~al.}(2024{\natexlab{a}})\citenamefont {Zhang}, \citenamefont {Lin},
  \citenamefont {Moreo}, \citenamefont {Maier},\ and\ \citenamefont
  {Dagotto}}]{zhangStructuralPhaseTransition2024}%
  \BibitemOpen
  \bibfield  {author} {\bibinfo {author} {\bibfnamefont {Y.}~\bibnamefont
  {Zhang}}, \bibinfo {author} {\bibfnamefont {L.-F.}\ \bibnamefont {Lin}},
  \bibinfo {author} {\bibfnamefont {A.}~\bibnamefont {Moreo}}, \bibinfo
  {author} {\bibfnamefont {T.~A.}\ \bibnamefont {Maier}},\ and\ \bibinfo
  {author} {\bibfnamefont {E.}~\bibnamefont {Dagotto}},\ }\bibfield  {title}
  {\bibinfo {title} {Structural phase transition, s{\textpm}-wave pairing, and
  magnetic stripe order in bilayered superconductor {{La3Ni2O7}} under
  pressure},\ }\href {https://doi.org/10.1038/s41467-024-46622-z} {\bibfield
  {journal} {\bibinfo  {journal} {Nature Communications}\ }\textbf {\bibinfo
  {volume} {15}},\ \bibinfo {pages} {2470} (\bibinfo {year}
  {2024}{\natexlab{a}})}\BibitemShut {NoStop}%
\bibitem [{\citenamefont {Lechermann}\ \emph {et~al.}(2023)\citenamefont
  {Lechermann}, \citenamefont {Gondolf}, \citenamefont {B{\"o}tzel},\ and\
  \citenamefont
  {Eremin}}]{lechermannElectronicCorrelationsSuperconducting2023}%
  \BibitemOpen
  \bibfield  {author} {\bibinfo {author} {\bibfnamefont {F.}~\bibnamefont
  {Lechermann}}, \bibinfo {author} {\bibfnamefont {J.}~\bibnamefont {Gondolf}},
  \bibinfo {author} {\bibfnamefont {S.}~\bibnamefont {B{\"o}tzel}},\ and\
  \bibinfo {author} {\bibfnamefont {I.~M.}\ \bibnamefont {Eremin}},\ }\bibfield
   {title} {\bibinfo {title} {Electronic correlations and superconducting
  instability in ${\mathrm{la}}_{3}{\mathrm{ni}}_{2}{\mathrm{o}}_{7}$ under
  high pressure},\ }\href {https://doi.org/10.1103/PhysRevB.108.L201121}
  {\bibfield  {journal} {\bibinfo  {journal} {Physical Review B}\ }\textbf
  {\bibinfo {volume} {108}},\ \bibinfo {pages} {L201121} (\bibinfo {year}
  {2023})}\BibitemShut {NoStop}%
\bibitem [{\citenamefont {Luo}\ \emph {et~al.}(2024)\citenamefont {Luo},
  \citenamefont {Lv}, \citenamefont {Wang}, \citenamefont {W{\'u}},\ and\
  \citenamefont {Yao}}]{luoHighTCSuperconductivityLa3Ni2O72024}%
  \BibitemOpen
  \bibfield  {author} {\bibinfo {author} {\bibfnamefont {Z.}~\bibnamefont
  {Luo}}, \bibinfo {author} {\bibfnamefont {B.}~\bibnamefont {Lv}}, \bibinfo
  {author} {\bibfnamefont {M.}~\bibnamefont {Wang}}, \bibinfo {author}
  {\bibfnamefont {W.}~\bibnamefont {W{\'u}}},\ and\ \bibinfo {author}
  {\bibfnamefont {D.-X.}\ \bibnamefont {Yao}},\ }\bibfield  {title} {\bibinfo
  {title} {High-{$T_c$} superconductivity in
  {${\mathrm{La}}_{3}{\mathrm{Ni}}_{2}{\mathrm{O}}_{7}$} based on the bilayer
  two-orbital t-{{J}} model},\ }\href
  {https://doi.org/10.1038/s41535-024-00668-w} {\bibfield  {journal} {\bibinfo
  {journal} {npj Quantum Materials}\ }\textbf {\bibinfo {volume} {9}},\
  \bibinfo {pages} {1} (\bibinfo {year} {2024})}\BibitemShut {NoStop}%
\bibitem [{\citenamefont {Yang}\ \emph {et~al.}(2023)\citenamefont {Yang},
  \citenamefont {Wang},\ and\ \citenamefont
  {Wang}}]{yangPossiblewaveSuperconductivity2023}%
  \BibitemOpen
  \bibfield  {author} {\bibinfo {author} {\bibfnamefont {Q.-G.}\ \bibnamefont
  {Yang}}, \bibinfo {author} {\bibfnamefont {D.}~\bibnamefont {Wang}},\ and\
  \bibinfo {author} {\bibfnamefont {Q.-H.}\ \bibnamefont {Wang}},\ }\bibfield
  {title} {\bibinfo {title} {Possible ${s}_{\ifmmode\pm\else\textpm\fi{}}$-wave
  superconductivity in ${\mathrm{la}}_{3}{\mathrm{ni}}_{2}{\mathrm{o}}_{7}$},\
  }\href {https://doi.org/10.1103/PhysRevB.108.L140505} {\bibfield  {journal}
  {\bibinfo  {journal} {Physical Review B}\ }\textbf {\bibinfo {volume}
  {108}},\ \bibinfo {pages} {L140505} (\bibinfo {year} {2023})}\BibitemShut
  {NoStop}%
\bibitem [{\citenamefont {Liu}\ \emph {et~al.}(2023)\citenamefont {Liu},
  \citenamefont {Mei}, \citenamefont {Ye}, \citenamefont {Chen},\ and\
  \citenamefont {Yang}}]{liuWavePairingDestructive2023}%
  \BibitemOpen
  \bibfield  {author} {\bibinfo {author} {\bibfnamefont {Y.-B.}\ \bibnamefont
  {Liu}}, \bibinfo {author} {\bibfnamefont {J.-W.}\ \bibnamefont {Mei}},
  \bibinfo {author} {\bibfnamefont {F.}~\bibnamefont {Ye}}, \bibinfo {author}
  {\bibfnamefont {W.-Q.}\ \bibnamefont {Chen}},\ and\ \bibinfo {author}
  {\bibfnamefont {F.}~\bibnamefont {Yang}},\ }\bibfield  {title} {\bibinfo
  {title} {${s}^{\ifmmode\pm\else\textpm\fi{}}$-wave pairing and the
  destructive role of apical-oxygen deficiencies in
  ${\mathrm{la}}_{3}{\mathrm{ni}}_{2}{\mathrm{o}}_{7}$ under pressure},\ }\href
  {https://doi.org/10.1103/PhysRevLett.131.236002} {\bibfield  {journal}
  {\bibinfo  {journal} {Physical Review Letters}\ }\textbf {\bibinfo {volume}
  {131}},\ \bibinfo {pages} {236002} (\bibinfo {year} {2023})}\BibitemShut
  {NoStop}%
\bibitem [{\citenamefont {Lechermann}\ \emph {et~al.}(2024)\citenamefont
  {Lechermann}, \citenamefont {B{\"o}tzel},\ and\ \citenamefont
  {Eremin}}]{lechermannElectronicInstabilityLayer2024}%
  \BibitemOpen
  \bibfield  {author} {\bibinfo {author} {\bibfnamefont {F.}~\bibnamefont
  {Lechermann}}, \bibinfo {author} {\bibfnamefont {S.}~\bibnamefont
  {B{\"o}tzel}},\ and\ \bibinfo {author} {\bibfnamefont {I.~M.}\ \bibnamefont
  {Eremin}},\ }\bibfield  {title} {\bibinfo {title} {Electronic instability,
  layer selectivity, and fermi arcs in
  ${\text{la}}_{3}{\text{ni}}_{2}{\text{o}}_{7}$},\ }\href
  {https://doi.org/10.1103/PhysRevMaterials.8.074802} {\bibfield  {journal}
  {\bibinfo  {journal} {Physical Review Materials}\ }\textbf {\bibinfo {volume}
  {8}},\ \bibinfo {pages} {074802} (\bibinfo {year} {2024})}\BibitemShut
  {NoStop}%
\bibitem [{\citenamefont {Qu}\ \emph {et~al.}(2024)\citenamefont {Qu},
  \citenamefont {Qu}, \citenamefont {Chen}, \citenamefont {Wu}, \citenamefont
  {Yang}, \citenamefont {Li},\ and\ \citenamefont {Su}}]{quBilayerModel2024}%
  \BibitemOpen
  \bibfield  {author} {\bibinfo {author} {\bibfnamefont {X.-Z.}\ \bibnamefont
  {Qu}}, \bibinfo {author} {\bibfnamefont {D.-W.}\ \bibnamefont {Qu}}, \bibinfo
  {author} {\bibfnamefont {J.}~\bibnamefont {Chen}}, \bibinfo {author}
  {\bibfnamefont {C.}~\bibnamefont {Wu}}, \bibinfo {author} {\bibfnamefont
  {F.}~\bibnamefont {Yang}}, \bibinfo {author} {\bibfnamefont {W.}~\bibnamefont
  {Li}},\ and\ \bibinfo {author} {\bibfnamefont {G.}~\bibnamefont {Su}},\
  }\bibfield  {title} {\bibinfo {title} {Bilayer
  ${t\text{\ensuremath{-}}J\text{\ensuremath{-}}J}_{\ensuremath{\perp}}$ model
  and magnetically mediated pairing in the pressurized nickelate
  ${\mathrm{la}}_{3}{\mathrm{ni}}_{2}{\mathrm{o}}_{7}$},\ }\href
  {https://doi.org/10.1103/PhysRevLett.132.036502} {\bibfield  {journal}
  {\bibinfo  {journal} {Physical Review Letters}\ }\textbf {\bibinfo {volume}
  {132}},\ \bibinfo {pages} {036502} (\bibinfo {year} {2024})}\BibitemShut
  {NoStop}%
\bibitem [{\citenamefont {Chen}\ \emph
  {et~al.}(2024{\natexlab{a}})\citenamefont {Chen}, \citenamefont {Yang},\ and\
  \citenamefont {Li}}]{chenOrbitalselectiveSuperconductivityPressurized2024}%
  \BibitemOpen
  \bibfield  {author} {\bibinfo {author} {\bibfnamefont {J.}~\bibnamefont
  {Chen}}, \bibinfo {author} {\bibfnamefont {F.}~\bibnamefont {Yang}},\ and\
  \bibinfo {author} {\bibfnamefont {W.}~\bibnamefont {Li}},\ }\bibfield
  {title} {\bibinfo {title} {Orbital-selective superconductivity in the
  pressurized bilayer nickelate
  ${\mathrm{la}}_{3}{\mathrm{ni}}_{2}{\mathrm{o}}_{7}$: An infinite projected
  entangled-pair state study},\ }\href
  {https://doi.org/10.1103/PhysRevB.110.L041111} {\bibfield  {journal}
  {\bibinfo  {journal} {Physical Review B}\ }\textbf {\bibinfo {volume}
  {110}},\ \bibinfo {pages} {L041111} (\bibinfo {year}
  {2024}{\natexlab{a}})}\BibitemShut {NoStop}%
\bibitem [{\citenamefont {Sakakibara}\ \emph {et~al.}(2024)\citenamefont
  {Sakakibara}, \citenamefont {Kitamine}, \citenamefont {Ochi},\ and\
  \citenamefont {Kuroki}}]{sakakibaraPossibleHigh$T_c$2024}%
  \BibitemOpen
  \bibfield  {author} {\bibinfo {author} {\bibfnamefont {H.}~\bibnamefont
  {Sakakibara}}, \bibinfo {author} {\bibfnamefont {N.}~\bibnamefont
  {Kitamine}}, \bibinfo {author} {\bibfnamefont {M.}~\bibnamefont {Ochi}},\
  and\ \bibinfo {author} {\bibfnamefont {K.}~\bibnamefont {Kuroki}},\
  }\bibfield  {title} {\bibinfo {title} {Possible high ${T}_{c}$
  superconductivity in ${\mathrm{la}}_{3}{\mathrm{ni}}_{2}{\mathrm{o}}_{7}$
  under high pressure through manifestation of a nearly half-filled bilayer
  hubbard model},\ }\href {https://doi.org/10.1103/PhysRevLett.132.106002}
  {\bibfield  {journal} {\bibinfo  {journal} {Physical Review Letters}\
  }\textbf {\bibinfo {volume} {132}},\ \bibinfo {pages} {106002} (\bibinfo
  {year} {2024})}\BibitemShut {NoStop}%
\bibitem [{\citenamefont {Fan}\ \emph {et~al.}(2024)\citenamefont {Fan},
  \citenamefont {Zhang}, \citenamefont {Zhan}, \citenamefont {Lv},
  \citenamefont {Jiang}, \citenamefont {Normand},\ and\ \citenamefont
  {Xiang}}]{fanSuperconductivityNickelateCuprate2024}%
  \BibitemOpen
  \bibfield  {author} {\bibinfo {author} {\bibfnamefont {Z.}~\bibnamefont
  {Fan}}, \bibinfo {author} {\bibfnamefont {J.-F.}\ \bibnamefont {Zhang}},
  \bibinfo {author} {\bibfnamefont {B.}~\bibnamefont {Zhan}}, \bibinfo {author}
  {\bibfnamefont {D.}~\bibnamefont {Lv}}, \bibinfo {author} {\bibfnamefont
  {X.-Y.}\ \bibnamefont {Jiang}}, \bibinfo {author} {\bibfnamefont
  {B.}~\bibnamefont {Normand}},\ and\ \bibinfo {author} {\bibfnamefont
  {T.}~\bibnamefont {Xiang}},\ }\bibfield  {title} {\bibinfo {title}
  {Superconductivity in nickelate and cuprate superconductors with strong
  bilayer coupling},\ }\href {https://doi.org/10.1103/PhysRevB.110.024514}
  {\bibfield  {journal} {\bibinfo  {journal} {Physical Review B}\ }\textbf
  {\bibinfo {volume} {110}},\ \bibinfo {pages} {024514} (\bibinfo {year}
  {2024})}\BibitemShut {NoStop}%
\bibitem [{\citenamefont {B{\"o}tzel}\ \emph {et~al.}(2024)\citenamefont
  {B{\"o}tzel}, \citenamefont {Lechermann}, \citenamefont {Gondolf},\ and\
  \citenamefont {Eremin}}]{botzelTheoryMagneticExcitations2024}%
  \BibitemOpen
  \bibfield  {author} {\bibinfo {author} {\bibfnamefont {S.}~\bibnamefont
  {B{\"o}tzel}}, \bibinfo {author} {\bibfnamefont {F.}~\bibnamefont
  {Lechermann}}, \bibinfo {author} {\bibfnamefont {J.}~\bibnamefont
  {Gondolf}},\ and\ \bibinfo {author} {\bibfnamefont {I.~M.}\ \bibnamefont
  {Eremin}},\ }\bibfield  {title} {\bibinfo {title} {Theory of magnetic
  excitations in the multilayer nickelate superconductor
  ${\mathrm{la}}_{3}{\mathrm{ni}}_{2}{\mathrm{o}}_{7}$},\ }\href
  {https://doi.org/10.1103/PhysRevB.109.L180502} {\bibfield  {journal}
  {\bibinfo  {journal} {Physical Review B}\ }\textbf {\bibinfo {volume}
  {109}},\ \bibinfo {pages} {L180502} (\bibinfo {year} {2024})}\BibitemShut
  {NoStop}%
\bibitem [{\citenamefont {Xia}\ \emph {et~al.}(2025)\citenamefont {Xia},
  \citenamefont {Liu}, \citenamefont {Zhou},\ and\ \citenamefont
  {Chen}}]{Xia:NatureCommunications:1054:2025}%
  \BibitemOpen
  \bibfield  {author} {\bibinfo {author} {\bibfnamefont {C.}~\bibnamefont
  {Xia}}, \bibinfo {author} {\bibfnamefont {H.}~\bibnamefont {Liu}}, \bibinfo
  {author} {\bibfnamefont {S.}~\bibnamefont {Zhou}},\ and\ \bibinfo {author}
  {\bibfnamefont {H.}~\bibnamefont {Chen}},\ }\bibfield  {title} {\bibinfo
  {title} {Sensitive dependence of pairing symmetry on ni-eg crystal field
  splitting in the nickelate superconductor la3ni2o7},\ }\href
  {https://doi.org/10.1038/s41467-025-56206-0} {\bibfield  {journal} {\bibinfo
  {journal} {Nature Communications}\ }\textbf {\bibinfo {volume} {16}},\
  \bibinfo {pages} {1054} (\bibinfo {year} {2025})}\BibitemShut {NoStop}%
\bibitem [{\citenamefont {Xie}\ \emph {et~al.}(2024)\citenamefont {Xie},
  \citenamefont {Huo}, \citenamefont {Ni}, \citenamefont {Shen}, \citenamefont
  {Huang}, \citenamefont {Sun}, \citenamefont {Walker}, \citenamefont {Adroja},
  \citenamefont {Yu}, \citenamefont {Shen}, \citenamefont {He}, \citenamefont
  {Cao},\ and\ \citenamefont {Wang}}]{xieStrongInterlayerMagnetic2024}%
  \BibitemOpen
  \bibfield  {author} {\bibinfo {author} {\bibfnamefont {T.}~\bibnamefont
  {Xie}}, \bibinfo {author} {\bibfnamefont {M.}~\bibnamefont {Huo}}, \bibinfo
  {author} {\bibfnamefont {X.}~\bibnamefont {Ni}}, \bibinfo {author}
  {\bibfnamefont {F.}~\bibnamefont {Shen}}, \bibinfo {author} {\bibfnamefont
  {X.}~\bibnamefont {Huang}}, \bibinfo {author} {\bibfnamefont
  {H.}~\bibnamefont {Sun}}, \bibinfo {author} {\bibfnamefont {H.~C.}\
  \bibnamefont {Walker}}, \bibinfo {author} {\bibfnamefont {D.}~\bibnamefont
  {Adroja}}, \bibinfo {author} {\bibfnamefont {D.}~\bibnamefont {Yu}}, \bibinfo
  {author} {\bibfnamefont {B.}~\bibnamefont {Shen}}, \bibinfo {author}
  {\bibfnamefont {L.}~\bibnamefont {He}}, \bibinfo {author} {\bibfnamefont
  {K.}~\bibnamefont {Cao}},\ and\ \bibinfo {author} {\bibfnamefont
  {M.}~\bibnamefont {Wang}},\ }\bibfield  {title} {\bibinfo {title} {Strong
  interlayer magnetic exchange coupling in {{La3Ni2O7}}-{$\delta$} revealed by
  inelastic neutron scattering},\ }\bibfield  {journal} {\bibinfo  {journal}
  {Science Bulletin}\ }\href {https://doi.org/10.1016/j.scib.2024.07.030}
  {10.1016/j.scib.2024.07.030} (\bibinfo {year} {2024})\BibitemShut {NoStop}%
\bibitem [{\citenamefont {Chen}\ \emph
  {et~al.}(2024{\natexlab{b}})\citenamefont {Chen}, \citenamefont {Choi},
  \citenamefont {Jiang}, \citenamefont {Mei}, \citenamefont {Jiang},
  \citenamefont {Li}, \citenamefont {Agrestini}, \citenamefont
  {Garcia-Fernandez}, \citenamefont {Sun}, \citenamefont {Huang}, \citenamefont
  {Shen}, \citenamefont {Wang}, \citenamefont {Hu}, \citenamefont {Lu},
  \citenamefont {Zhou},\ and\ \citenamefont
  {Feng}}]{Chen:NatureCommunications:9597:2024}%
  \BibitemOpen
  \bibfield  {author} {\bibinfo {author} {\bibfnamefont {X.}~\bibnamefont
  {Chen}}, \bibinfo {author} {\bibfnamefont {J.}~\bibnamefont {Choi}}, \bibinfo
  {author} {\bibfnamefont {Z.}~\bibnamefont {Jiang}}, \bibinfo {author}
  {\bibfnamefont {J.}~\bibnamefont {Mei}}, \bibinfo {author} {\bibfnamefont
  {K.}~\bibnamefont {Jiang}}, \bibinfo {author} {\bibfnamefont
  {J.}~\bibnamefont {Li}}, \bibinfo {author} {\bibfnamefont {S.}~\bibnamefont
  {Agrestini}}, \bibinfo {author} {\bibfnamefont {M.}~\bibnamefont
  {Garcia-Fernandez}}, \bibinfo {author} {\bibfnamefont {H.}~\bibnamefont
  {Sun}}, \bibinfo {author} {\bibfnamefont {X.}~\bibnamefont {Huang}}, \bibinfo
  {author} {\bibfnamefont {D.}~\bibnamefont {Shen}}, \bibinfo {author}
  {\bibfnamefont {M.}~\bibnamefont {Wang}}, \bibinfo {author} {\bibfnamefont
  {J.}~\bibnamefont {Hu}}, \bibinfo {author} {\bibfnamefont {Y.}~\bibnamefont
  {Lu}}, \bibinfo {author} {\bibfnamefont {K.-J.}\ \bibnamefont {Zhou}},\ and\
  \bibinfo {author} {\bibfnamefont {D.}~\bibnamefont {Feng}},\ }\bibfield
  {title} {\bibinfo {title} {Electronic and magnetic excitations in la3ni2o7},\
  }\href {https://doi.org/10.1038/s41467-024-53863-5} {\bibfield  {journal}
  {\bibinfo  {journal} {Nature Communications}\ }\textbf {\bibinfo {volume}
  {15}},\ \bibinfo {pages} {9597} (\bibinfo {year}
  {2024}{\natexlab{b}})}\BibitemShut {NoStop}%
\bibitem [{\citenamefont {Zhong}\ \emph {et~al.}(2025)\citenamefont {Zhong},
  \citenamefont {Hao}, \citenamefont {Wei}, \citenamefont {Zhang},
  \citenamefont {Liu}, \citenamefont {Huang}, \citenamefont {Ni}, \citenamefont
  {dos Reis~Cantarino}, \citenamefont {Cao}, \citenamefont {Nie}, \citenamefont
  {Schmitt},\ and\ \citenamefont {Lu}}]{Zhong_arXiv.org2502.03178__2025}%
  \BibitemOpen
  \bibfield  {author} {\bibinfo {author} {\bibfnamefont {H.}~\bibnamefont
  {Zhong}}, \bibinfo {author} {\bibfnamefont {B.}~\bibnamefont {Hao}}, \bibinfo
  {author} {\bibfnamefont {Y.}~\bibnamefont {Wei}}, \bibinfo {author}
  {\bibfnamefont {Z.}~\bibnamefont {Zhang}}, \bibinfo {author} {\bibfnamefont
  {R.}~\bibnamefont {Liu}}, \bibinfo {author} {\bibfnamefont {X.}~\bibnamefont
  {Huang}}, \bibinfo {author} {\bibfnamefont {X.-S.}\ \bibnamefont {Ni}},
  \bibinfo {author} {\bibfnamefont {M.}~\bibnamefont {dos Reis~Cantarino}},
  \bibinfo {author} {\bibfnamefont {K.}~\bibnamefont {Cao}}, \bibinfo {author}
  {\bibfnamefont {Y.}~\bibnamefont {Nie}}, \bibinfo {author} {\bibfnamefont
  {T.}~\bibnamefont {Schmitt}},\ and\ \bibinfo {author} {\bibfnamefont
  {X.}~\bibnamefont {Lu}},\ }\bibfield  {title} {\bibinfo {title} {Epitaxial
  strain tuning of electronic and spin excitations in la$_3$ni$_2$o$_7$ thin
  films},\ }\href {https://arxiv.org/abs/2502.03178} {\bibfield  {journal}
  {\bibinfo  {journal} {arXiv.org:2502.03178}\ } (\bibinfo {year}
  {2025})}\BibitemShut {NoStop}%
\bibitem [{\citenamefont {Graser}\ \emph {et~al.}(2009)\citenamefont {Graser},
  \citenamefont {Maier}, \citenamefont {Hirschfeld},\ and\ \citenamefont
  {Scalapino}}]{graserDegeneracySeveralPairing2009}%
  \BibitemOpen
  \bibfield  {author} {\bibinfo {author} {\bibfnamefont {S.}~\bibnamefont
  {Graser}}, \bibinfo {author} {\bibfnamefont {T.~A.}\ \bibnamefont {Maier}},
  \bibinfo {author} {\bibfnamefont {P.~J.}\ \bibnamefont {Hirschfeld}},\ and\
  \bibinfo {author} {\bibfnamefont {D.~J.}\ \bibnamefont {Scalapino}},\
  }\bibfield  {title} {\bibinfo {title} {Near-degeneracy of several pairing
  channels in multiorbital models for the {{Fe}} pnictides},\ }\href
  {https://doi.org/10.1088/1367-2630/11/2/025016} {\bibfield  {journal}
  {\bibinfo  {journal} {New Journal of Physics}\ }\textbf {\bibinfo {volume}
  {11}},\ \bibinfo {pages} {025016} (\bibinfo {year} {2009})}\BibitemShut
  {NoStop}%
\bibitem [{\citenamefont {Kemper}\ \emph {et~al.}(2010)\citenamefont {Kemper},
  \citenamefont {Maier}, \citenamefont {Graser}, \citenamefont {Cheng},
  \citenamefont {Hirschfeld},\ and\ \citenamefont
  {Scalapino}}]{kemperSensitivitySuperconductingState2010}%
  \BibitemOpen
  \bibfield  {author} {\bibinfo {author} {\bibfnamefont {A.~F.}\ \bibnamefont
  {Kemper}}, \bibinfo {author} {\bibfnamefont {T.~A.}\ \bibnamefont {Maier}},
  \bibinfo {author} {\bibfnamefont {S.}~\bibnamefont {Graser}}, \bibinfo
  {author} {\bibfnamefont {H.-P.}\ \bibnamefont {Cheng}}, \bibinfo {author}
  {\bibfnamefont {P.~J.}\ \bibnamefont {Hirschfeld}},\ and\ \bibinfo {author}
  {\bibfnamefont {D.~J.}\ \bibnamefont {Scalapino}},\ }\bibfield  {title}
  {\bibinfo {title} {Sensitivity of the superconducting state and magnetic
  susceptibility to key aspects of electronic structure in ferropnictides},\
  }\href {https://doi.org/10.1088/1367-2630/12/7/073030} {\bibfield  {journal}
  {\bibinfo  {journal} {New Journal of Physics}\ }\textbf {\bibinfo {volume}
  {12}},\ \bibinfo {pages} {073030} (\bibinfo {year} {2010})}\BibitemShut
  {NoStop}%
\bibitem [{\citenamefont {Altmeyer}\ \emph {et~al.}(2016)\citenamefont
  {Altmeyer}, \citenamefont {Guterding}, \citenamefont {Hirschfeld},
  \citenamefont {Maier}, \citenamefont {Valent{\'i}},\ and\ \citenamefont
  {Scalapino}}]{altmeyerRoleVertexCorrections2016}%
  \BibitemOpen
  \bibfield  {author} {\bibinfo {author} {\bibfnamefont {M.}~\bibnamefont
  {Altmeyer}}, \bibinfo {author} {\bibfnamefont {D.}~\bibnamefont {Guterding}},
  \bibinfo {author} {\bibfnamefont {P.~J.}\ \bibnamefont {Hirschfeld}},
  \bibinfo {author} {\bibfnamefont {T.~A.}\ \bibnamefont {Maier}}, \bibinfo
  {author} {\bibfnamefont {R.}~\bibnamefont {Valent{\'i}}},\ and\ \bibinfo
  {author} {\bibfnamefont {D.~J.}\ \bibnamefont {Scalapino}},\ }\bibfield
  {title} {\bibinfo {title} {Role of vertex corrections in the matrix
  formulation of the random phase approximation for the multiorbital
  {{Hubbard}} model},\ }\href {https://doi.org/10.1103/PhysRevB.94.214515}
  {\bibfield  {journal} {\bibinfo  {journal} {Physical Review B}\ }\textbf
  {\bibinfo {volume} {94}},\ \bibinfo {pages} {214515} (\bibinfo {year}
  {2016})}\BibitemShut {NoStop}%
\bibitem [{\citenamefont {Martins}\ \emph {et~al.}(2013)\citenamefont
  {Martins}, \citenamefont {Moreo},\ and\ \citenamefont
  {Dagotto}}]{Martins:Phys.Rev.B:081102:2013}%
  \BibitemOpen
  \bibfield  {author} {\bibinfo {author} {\bibfnamefont {G.~B.}\ \bibnamefont
  {Martins}}, \bibinfo {author} {\bibfnamefont {A.}~\bibnamefont {Moreo}},\
  and\ \bibinfo {author} {\bibfnamefont {E.}~\bibnamefont {Dagotto}},\
  }\bibfield  {title} {\bibinfo {title} {Rpa analysis of a two-orbital model
  for the bis${}_{2}$-based superconductors},\ }\href
  {https://doi.org/10.1103/PhysRevB.87.081102} {\bibfield  {journal} {\bibinfo
  {journal} {Phys. Rev. B}\ }\textbf {\bibinfo {volume} {87}},\ \bibinfo
  {pages} {081102} (\bibinfo {year} {2013})}\BibitemShut {NoStop}%
\bibitem [{\citenamefont {Braz}\ \emph {et~al.}(2024)\citenamefont {Braz},
  \citenamefont {Martins},\ and\ \citenamefont {{da
  Silva}}}]{brazChargeSpinFluctuations2024}%
  \BibitemOpen
  \bibfield  {author} {\bibinfo {author} {\bibfnamefont {L.~B.}\ \bibnamefont
  {Braz}}, \bibinfo {author} {\bibfnamefont {G.~B.}\ \bibnamefont {Martins}},\
  and\ \bibinfo {author} {\bibfnamefont {L.~G. G. V.~D.}\ \bibnamefont {{da
  Silva}}},\ }\href@noop {} {\bibinfo {title} {Charge and spin fluctuations in
  superconductors with intersublattice and interorbital interactions}},\
  \bibinfo {howpublished} {https://arxiv.org/abs/2403.02453v1} (\bibinfo {year}
  {2024})\BibitemShut {NoStop}%
\bibitem [{\citenamefont {Cai}\ \emph {et~al.}(2024)\citenamefont {Cai},
  \citenamefont {Zhou}, \citenamefont {Sun}, \citenamefont {Zhang},
  \citenamefont {Zhao}, \citenamefont {Huo}, \citenamefont {Nataf},
  \citenamefont {Wang}, \citenamefont {Li}, \citenamefont {Guo}, \citenamefont
  {Jiang}, \citenamefont {Wang}, \citenamefont {Ding}, \citenamefont {Yang},
  \citenamefont {Lu}, \citenamefont {Kong}, \citenamefont {Wu}, \citenamefont
  {Hu}, \citenamefont {Xiang}, \citenamefont {kwang Mao},\ and\ \citenamefont
  {Sun}}]{Cai:arXiv:2412.18343:2024}%
  \BibitemOpen
  \bibfield  {author} {\bibinfo {author} {\bibfnamefont {S.}~\bibnamefont
  {Cai}}, \bibinfo {author} {\bibfnamefont {Y.}~\bibnamefont {Zhou}}, \bibinfo
  {author} {\bibfnamefont {H.}~\bibnamefont {Sun}}, \bibinfo {author}
  {\bibfnamefont {K.}~\bibnamefont {Zhang}}, \bibinfo {author} {\bibfnamefont
  {J.}~\bibnamefont {Zhao}}, \bibinfo {author} {\bibfnamefont {M.}~\bibnamefont
  {Huo}}, \bibinfo {author} {\bibfnamefont {L.}~\bibnamefont {Nataf}}, \bibinfo
  {author} {\bibfnamefont {Y.}~\bibnamefont {Wang}}, \bibinfo {author}
  {\bibfnamefont {J.}~\bibnamefont {Li}}, \bibinfo {author} {\bibfnamefont
  {J.}~\bibnamefont {Guo}}, \bibinfo {author} {\bibfnamefont {K.}~\bibnamefont
  {Jiang}}, \bibinfo {author} {\bibfnamefont {M.}~\bibnamefont {Wang}},
  \bibinfo {author} {\bibfnamefont {Y.}~\bibnamefont {Ding}}, \bibinfo {author}
  {\bibfnamefont {W.}~\bibnamefont {Yang}}, \bibinfo {author} {\bibfnamefont
  {Y.}~\bibnamefont {Lu}}, \bibinfo {author} {\bibfnamefont {Q.}~\bibnamefont
  {Kong}}, \bibinfo {author} {\bibfnamefont {Q.}~\bibnamefont {Wu}}, \bibinfo
  {author} {\bibfnamefont {J.}~\bibnamefont {Hu}}, \bibinfo {author}
  {\bibfnamefont {T.}~\bibnamefont {Xiang}}, \bibinfo {author} {\bibfnamefont
  {H.}~\bibnamefont {kwang Mao}},\ and\ \bibinfo {author} {\bibfnamefont
  {L.}~\bibnamefont {Sun}},\ }\bibfield  {title} {\bibinfo {title}
  {Low-temperature mean valence of nickel ions in pressurized
  la$_3$ni$_2$o$_7$},\ }\href {https://arxiv.org/abs/2412.18343} {\bibfield
  {journal} {\bibinfo  {journal} {arXiv.org:2412.18343}\ } (\bibinfo {year}
  {2024})}\BibitemShut {NoStop}%
\bibitem [{\citenamefont {Yang}\ \emph {et~al.}(2024)\citenamefont {Yang},
  \citenamefont {Sun}, \citenamefont {Hu}, \citenamefont {Xie}, \citenamefont
  {Miao}, \citenamefont {Luo}, \citenamefont {Chen}, \citenamefont {Liang},
  \citenamefont {Zhu}, \citenamefont {Qu}, \citenamefont {Chen}, \citenamefont
  {Huo}, \citenamefont {Huang}, \citenamefont {Zhang}, \citenamefont {Zhang},
  \citenamefont {Yang}, \citenamefont {Wang}, \citenamefont {Peng},
  \citenamefont {Mao}, \citenamefont {Liu}, \citenamefont {Xu}, \citenamefont
  {Qian}, \citenamefont {Yao}, \citenamefont {Wang}, \citenamefont {Zhao},\
  and\ \citenamefont {Zhou}}]{yangOrbitaldependentElectronCorrelation2024}%
  \BibitemOpen
  \bibfield  {author} {\bibinfo {author} {\bibfnamefont {J.}~\bibnamefont
  {Yang}}, \bibinfo {author} {\bibfnamefont {H.}~\bibnamefont {Sun}}, \bibinfo
  {author} {\bibfnamefont {X.}~\bibnamefont {Hu}}, \bibinfo {author}
  {\bibfnamefont {Y.}~\bibnamefont {Xie}}, \bibinfo {author} {\bibfnamefont
  {T.}~\bibnamefont {Miao}}, \bibinfo {author} {\bibfnamefont {H.}~\bibnamefont
  {Luo}}, \bibinfo {author} {\bibfnamefont {H.}~\bibnamefont {Chen}}, \bibinfo
  {author} {\bibfnamefont {B.}~\bibnamefont {Liang}}, \bibinfo {author}
  {\bibfnamefont {W.}~\bibnamefont {Zhu}}, \bibinfo {author} {\bibfnamefont
  {G.}~\bibnamefont {Qu}}, \bibinfo {author} {\bibfnamefont {C.-Q.}\
  \bibnamefont {Chen}}, \bibinfo {author} {\bibfnamefont {M.}~\bibnamefont
  {Huo}}, \bibinfo {author} {\bibfnamefont {Y.}~\bibnamefont {Huang}}, \bibinfo
  {author} {\bibfnamefont {S.}~\bibnamefont {Zhang}}, \bibinfo {author}
  {\bibfnamefont {F.}~\bibnamefont {Zhang}}, \bibinfo {author} {\bibfnamefont
  {F.}~\bibnamefont {Yang}}, \bibinfo {author} {\bibfnamefont {Z.}~\bibnamefont
  {Wang}}, \bibinfo {author} {\bibfnamefont {Q.}~\bibnamefont {Peng}}, \bibinfo
  {author} {\bibfnamefont {H.}~\bibnamefont {Mao}}, \bibinfo {author}
  {\bibfnamefont {G.}~\bibnamefont {Liu}}, \bibinfo {author} {\bibfnamefont
  {Z.}~\bibnamefont {Xu}}, \bibinfo {author} {\bibfnamefont {T.}~\bibnamefont
  {Qian}}, \bibinfo {author} {\bibfnamefont {D.-X.}\ \bibnamefont {Yao}},
  \bibinfo {author} {\bibfnamefont {M.}~\bibnamefont {Wang}}, \bibinfo {author}
  {\bibfnamefont {L.}~\bibnamefont {Zhao}},\ and\ \bibinfo {author}
  {\bibfnamefont {X.~J.}\ \bibnamefont {Zhou}},\ }\bibfield  {title} {\bibinfo
  {title} {Orbital-dependent electron correlation in double-layer nickelate
  {{La3Ni2O7}}},\ }\href {https://doi.org/10.1038/s41467-024-48701-7}
  {\bibfield  {journal} {\bibinfo  {journal} {Nature Communications}\ }\textbf
  {\bibinfo {volume} {15}},\ \bibinfo {pages} {4373} (\bibinfo {year}
  {2024})}\BibitemShut {NoStop}%
\bibitem [{\citenamefont {Li}\ \emph {et~al.}(2024{\natexlab{b}})\citenamefont
  {Li}, \citenamefont {Du}, \citenamefont {Cao}, \citenamefont {Pei},
  \citenamefont {Zhang}, \citenamefont {Zhao}, \citenamefont {Zhai},
  \citenamefont {Xu}, \citenamefont {Liu}, \citenamefont {Li}, \citenamefont
  {Zhao}, \citenamefont {Li}, \citenamefont {Qi}, \citenamefont {Guo},
  \citenamefont {Chen},\ and\ \citenamefont
  {Yang}}]{liElectronicCorrelationPseudogapBehavior2024}%
  \BibitemOpen
  \bibfield  {author} {\bibinfo {author} {\bibfnamefont {Y.}~\bibnamefont
  {Li}}, \bibinfo {author} {\bibfnamefont {X.}~\bibnamefont {Du}}, \bibinfo
  {author} {\bibfnamefont {Y.}~\bibnamefont {Cao}}, \bibinfo {author}
  {\bibfnamefont {C.}~\bibnamefont {Pei}}, \bibinfo {author} {\bibfnamefont
  {M.}~\bibnamefont {Zhang}}, \bibinfo {author} {\bibfnamefont
  {W.}~\bibnamefont {Zhao}}, \bibinfo {author} {\bibfnamefont {K.}~\bibnamefont
  {Zhai}}, \bibinfo {author} {\bibfnamefont {R.}~\bibnamefont {Xu}}, \bibinfo
  {author} {\bibfnamefont {Z.}~\bibnamefont {Liu}}, \bibinfo {author}
  {\bibfnamefont {Z.}~\bibnamefont {Li}}, \bibinfo {author} {\bibfnamefont
  {J.}~\bibnamefont {Zhao}}, \bibinfo {author} {\bibfnamefont {G.}~\bibnamefont
  {Li}}, \bibinfo {author} {\bibfnamefont {Y.}~\bibnamefont {Qi}}, \bibinfo
  {author} {\bibfnamefont {H.}~\bibnamefont {Guo}}, \bibinfo {author}
  {\bibfnamefont {Y.}~\bibnamefont {Chen}},\ and\ \bibinfo {author}
  {\bibfnamefont {L.}~\bibnamefont {Yang}},\ }\bibfield  {title} {\bibinfo
  {title} {Electronic {{Correlation}} and {{Pseudogap-Like Behavior}} of
  {{High-Temperature Superconductor La3Ni2O7}}},\ }\href
  {https://doi.org/10.1088/0256-307X/41/8/087402} {\bibfield  {journal}
  {\bibinfo  {journal} {Chinese Physics Letters}\ }\textbf {\bibinfo {volume}
  {41}},\ \bibinfo {pages} {087402} (\bibinfo {year}
  {2024}{\natexlab{b}})}\BibitemShut {NoStop}%
\bibitem [{\citenamefont {Zhang}\ \emph
  {et~al.}(2024{\natexlab{b}})\citenamefont {Zhang}, \citenamefont {Lin},
  \citenamefont {Moreo}, \citenamefont {Maier},\ and\ \citenamefont
  {Dagotto}}]{Zhang:Phys.Rev.B:045151:2024}%
  \BibitemOpen
  \bibfield  {author} {\bibinfo {author} {\bibfnamefont {Y.}~\bibnamefont
  {Zhang}}, \bibinfo {author} {\bibfnamefont {L.-F.}\ \bibnamefont {Lin}},
  \bibinfo {author} {\bibfnamefont {A.}~\bibnamefont {Moreo}}, \bibinfo
  {author} {\bibfnamefont {T.~A.}\ \bibnamefont {Maier}},\ and\ \bibinfo
  {author} {\bibfnamefont {E.}~\bibnamefont {Dagotto}},\ }\bibfield  {title}
  {\bibinfo {title} {Electronic structure, magnetic correlations, and
  superconducting pairing in the reduced ruddlesden-popper bilayer
  ${\mathrm{la}}_{3}{\mathrm{ni}}_{2}{\mathrm{o}}_{6}$ under pressure:
  Different role of ${d}_{3{z}^{2}\ensuremath{-}{r}^{2}}$ orbital compared with
  ${\mathrm{la}}_{3}{\mathrm{ni}}_{2}{\mathrm{o}}_{7}$},\ }\href
  {https://doi.org/10.1103/PhysRevB.109.045151} {\bibfield  {journal} {\bibinfo
   {journal} {Phys. Rev. B}\ }\textbf {\bibinfo {volume} {109}},\ \bibinfo
  {pages} {045151} (\bibinfo {year} {2024}{\natexlab{b}})}\BibitemShut
  {NoStop}%
\bibitem [{\citenamefont {Zhang}\ \emph
  {et~al.}(2024{\natexlab{c}})\citenamefont {Zhang}, \citenamefont {Lin},
  \citenamefont {Moreo}, \citenamefont {Maier},\ and\ \citenamefont
  {Dagotto}}]{Zhang:Phys.Rev.Lett.:136001:2024}%
  \BibitemOpen
  \bibfield  {author} {\bibinfo {author} {\bibfnamefont {Y.}~\bibnamefont
  {Zhang}}, \bibinfo {author} {\bibfnamefont {L.-F.}\ \bibnamefont {Lin}},
  \bibinfo {author} {\bibfnamefont {A.}~\bibnamefont {Moreo}}, \bibinfo
  {author} {\bibfnamefont {T.~A.}\ \bibnamefont {Maier}},\ and\ \bibinfo
  {author} {\bibfnamefont {E.}~\bibnamefont {Dagotto}},\ }\bibfield  {title}
  {\bibinfo {title} {Prediction of ${s}^{\pm}$-wave superconductivity enhanced
  by electronic doping in trilayer nickelates
  ${\mathrm{la}}_{4}{\mathrm{ni}}_{3}{\mathrm{o}}_{10}$ under pressure},\
  }\href {https://doi.org/10.1103/PhysRevLett.133.136001} {\bibfield  {journal}
  {\bibinfo  {journal} {Phys. Rev. Lett.}\ }\textbf {\bibinfo {volume} {133}},\
  \bibinfo {pages} {136001} (\bibinfo {year} {2024}{\natexlab{c}})}\BibitemShut
  {NoStop}%
\bibitem [{Note1()}]{Note1}%
  \BibitemOpen
  \bibinfo {note} {The remaining matrix elements can be trivially obtained by
  exchanging $A \leftrightarrow B$}\BibitemShut {NoStop}%
\bibitem [{\citenamefont
  {Ozaki}(2007)}]{ozakiContinuedFractionRepresentation2007}%
  \BibitemOpen
  \bibfield  {author} {\bibinfo {author} {\bibfnamefont {T.}~\bibnamefont
  {Ozaki}},\ }\bibfield  {title} {\bibinfo {title} {Continued fraction
  representation of the {{Fermi-Dirac}} function for large-scale electronic
  structure calculations},\ }\href {https://doi.org/10.1103/PhysRevB.75.035123}
  {\bibfield  {journal} {\bibinfo  {journal} {Physical Review B}\ }\textbf
  {\bibinfo {volume} {75}},\ \bibinfo {pages} {035123} (\bibinfo {year}
  {2007})}\BibitemShut {NoStop}%
\bibitem [{\citenamefont {Xi}\ \emph {et~al.}(2025)\citenamefont {Xi},
  \citenamefont {Yu},\ and\ \citenamefont {Li}}]{Xi_Phys.Rev.B_104505_2025}%
  \BibitemOpen
  \bibfield  {author} {\bibinfo {author} {\bibfnamefont {W.}~\bibnamefont
  {Xi}}, \bibinfo {author} {\bibfnamefont {S.-L.}\ \bibnamefont {Yu}},\ and\
  \bibinfo {author} {\bibfnamefont {J.-X.}\ \bibnamefont {Li}},\ }\bibfield
  {title} {\bibinfo {title} {Transition from ${s}_{pm}$-wave to
  ${d}_{{x}^{2}\ensuremath{-}{y}^{2}}$-wave superconductivity driven by
  interlayer interaction in the bilayer two-orbital model of
  ${\text{la}}_{3}{\text{ni}}_{2}{\text{o}}_{7}$},\ }\href
  {https://doi.org/10.1103/PhysRevB.111.104505} {\bibfield  {journal} {\bibinfo
   {journal} {Phys. Rev. B}\ }\textbf {\bibinfo {volume} {111}},\ \bibinfo
  {pages} {104505} (\bibinfo {year} {2025})}\BibitemShut {NoStop}%
\bibitem [{\citenamefont {Zhan}\ \emph {et~al.}(2025)\citenamefont {Zhan},
  \citenamefont {Le}, \citenamefont {Wu},\ and\ \citenamefont
  {Hu}}]{Zhan:arXiv:2503.18877:2025}%
  \BibitemOpen
  \bibfield  {author} {\bibinfo {author} {\bibfnamefont {J.}~\bibnamefont
  {Zhan}}, \bibinfo {author} {\bibfnamefont {C.}~\bibnamefont {Le}}, \bibinfo
  {author} {\bibfnamefont {X.}~\bibnamefont {Wu}},\ and\ \bibinfo {author}
  {\bibfnamefont {J.}~\bibnamefont {Hu}},\ }\bibfield  {title} {\bibinfo
  {title} {Impact of nonlocal coulomb repulsion on superconductivity and
  density-wave orders in bilayer nickelates},\ }\href
  {https://arxiv.org/abs/2503.18877} {\bibfield  {journal} {\bibinfo  {journal}
  {arXiv.org:2503.18877}\ } (\bibinfo {year} {2025})}\BibitemShut {NoStop}%
\bibitem [{\citenamefont {Barreto~Braz}\ \emph {et~al.}(2025)\citenamefont
  {Barreto~Braz}, \citenamefont {Dias~da Silva},\ and\ \citenamefont
  {Balster~Martins}}]{ZenodoData}%
  \BibitemOpen
  \bibfield  {author} {\bibinfo {author} {\bibfnamefont {L.}~\bibnamefont
  {Barreto~Braz}}, \bibinfo {author} {\bibfnamefont {L.}~\bibnamefont {Dias~da
  Silva}},\ and\ \bibinfo {author} {\bibfnamefont {G.}~\bibnamefont
  {Balster~Martins}},\ }\bibfield  {title} {\bibinfo {title} {Interlayer
  interactions in la$_3$ni$_2$o$_7$ under pressure: from $s^{\pm}$ to
  $d\_{xy}$-wave superconductivity},\ }\href
  {https://doi.org/10.5281/zenodo.14933119} {10.5281/zenodo.14933119} (\bibinfo
  {year} {2025})\BibitemShut {NoStop}%
\end{thebibliography}

%

\end{document}